\documentclass[final,3p,times,author,twocolumn]{elsarticle}

\usepackage[ruled,vlined]{algorithm2e}
\usepackage{amsmath,amssymb}
\usepackage{url}

\usepackage{graphicx}
\usepackage{tikz}
\usetikzlibrary{arrows.meta, positioning, shapes.geometric, fit, calc, decorations.pathmorphing, decorations.pathreplacing}
\usepackage{booktabs}
\usepackage{multirow}
\usepackage{enumitem}
\usepackage{xcolor}
\usepackage{colortbl}
\usepackage{lineno}
\usepackage{balance}
\usepackage{hyperref}

\journal{Computers \& Security}

\definecolor{accentblue}{HTML}{2563EB}
\definecolor{accentteal}{HTML}{0D9488}
\definecolor{accentred}{HTML}{DC2626}
\definecolor{accentamber}{HTML}{D97706}
\definecolor{accentgreen}{HTML}{16A34A}
\definecolor{softblue}{HTML}{DBEAFE}
\definecolor{softgreen}{HTML}{DCFCE7}
\definecolor{softred}{HTML}{FEE2E2}
\definecolor{softamber}{HTML}{FEF3C7}
\definecolor{softteal}{HTML}{CCFBF1}
\definecolor{darkgray}{HTML}{374151}
\definecolor{lightgray}{HTML}{F3F4F6}

\usepackage{mdframed}
\newmdenv[
  skipabove=6pt,
  skipbelow=6pt,
  linewidth=1pt,
  linecolor=accentblue,
  backgroundcolor=softblue,
  roundcorner=4pt
]{takeawaybox}

\newcommand{\tsc}[1]{\textsc{#1}}

\begin{document}

\begin{frontmatter}

%\title{The Double Dip: Reward Abuse via Refund Insensitive Business Logic}
\title{Refunded but Rewarded: The Double Dip Attack on Cashback Reward Engines}
\author[tulsa]{S~M~Zia~Ur~Rashid\corref{cor1}}
\ead{ziaur-rashid@utulsa.edu}
\author[tulsa]{Suman~Rath}
\ead{suman-rath@utulsa.edu}
\cortext[cor1]{Corresponding author}
\affiliation[tulsa]{organization={Department of Electrical \& Computer Engineering, The University of Tulsa},
            city={Tulsa},
            state={Oklahoma},
            country={USA}}

\begin{abstract}
Cashback and loyalty reward programs now serve as central instruments in the competitive landscape of cards, digital wallets, and payment platforms. Despite their financial significance, the business logic governing these programs is seldom treated as a security critical surface. In this paper, we study a class of \emph{reward abuse attacks} that arise from flaws in how reward systems accrue, redeem, and adjust incentives when underlying transactions are reversed through refunds. Using controlled, small scale experiments on six issuer accounts we legitimately hold, we document a spectrum of real world behaviors in production systems. At one extreme, a debit based cashback program (Issuer~A) never adjusts rewards when refunded transactions post, enabling a deterministic \emph{double dip cashback reward abuse attack}. A credit card program (Issuer~B) exhibits an analogous reward integrity violation through a statement cycle timing gap that allows reward redemption before the merchant return window closes. At an intermediate tier, a credit card issuer (Issuer~F) creates negative reward entries on refunds at statement close but makes rewards redeemable immediately upon settlement, creating a timing asymmetry that allows users to extract reward value before clawback occurs. At the robust end, three credit card issuers (C, D, and E) implement indefinite negative balance enforcement with proportional clawback. We formalize reward engines as state machines, introduce two integrity invariants (\emph{Reward Integrity} and \emph{Refund Reward Consistency}), develop a taxonomy of vulnerability classes mapped to CWE and OWASP categories, and present defensive pseudo algorithms with a semi formal correctness argument that close the identified loopholes. The primary vulnerability (Issuer~A) was reported through a private bug bounty program and has been acknowledged by the vendor; good faith disclosure efforts for Issuer~B are detailed in Section~\ref{sec:ethics}.
\end{abstract}

\begin{keyword}
cashback reward abuse \sep business logic vulnerability \sep double-dip attack \sep refund fraud \sep cashback reward engines \sep payment security
\end{keyword}

\end{frontmatter}

% ============================================================
% 1  INTRODUCTION
% ============================================================
\section{Introduction}
\label{sec:intro}

Every major credit card, debit card, and digital wallet in the United States now ships with some form of cashback or loyalty reward program. These programs collectively move billions of dollars in incentive value each year, and for many issuers they serve as a primary lever to acquire customers, steer spending, and generate interchange revenue~\cite{taleizadeh2023cashback,wu2023praise}. Yet while the payment authorization and settlement systems that underlie these programs are engineered with decades of security scrutiny, the reward engines that sit on top of them receive far less attention. In many organizations, the logic that decides when to grant, adjust, or revoke cashback lives closer to marketing teams than to security engineering.

This paper demonstrates that reward engines can harbor exploitable business logic vulnerabilities with direct financial consequences. We discovered and confirmed, on accounts we personally hold, that certain production cashback programs allow a user to earn rewards on a purchase, redeem those rewards for cash or statement credit, and then return the merchandise for a full refund without any corresponding adjustment to the reward ledger. The user ends the cycle with zero net spend and positive net reward, a deterministic profit that can be repeated every billing period up to the program's monthly cap.

We use the term \emph{reward abuse attack} to describe systematic, intentional exploitation of cashback and loyalty reward mechanisms to extract monetary benefit beyond the platform's intended design, while using legitimate accounts and nominally valid transactions. This definition distinguishes our concern from classic payment fraud involving stolen or cloned cards, from casual deal hunting that stays within published terms, and from friendly fraud or chargeback fraud where users dispute legitimate charges as unauthorized. Our focus is on logic flaws in how rewards are accounted for in response to refunds, not on identity theft or claim falsification.

Our work is grounded in a concrete pattern we call the \emph{double dip cashback reward abuse attack} (DDRA): a user earns and redeems cashback on a purchase, then later reverses the underlying spend via a refund without any proportional clawback of rewards. We discovered two distinct variants of this pattern operating in production financial systems, each exploiting different architectural choices in the reward engine.

We make four main contributions:

\begin{enumerate}[leftmargin=*]
  \item \textbf{Empirical case studies.}
  We instantiate a three tier vulnerability taxonomy using six production issuer accounts we legitimately hold:
  \begin{itemize}[leftmargin=*]
    \item \textbf{Case~I (Issuer~A):} A debit based cashback program that completely fails to adjust rewards when underlying transactions are refunded, enabling a deterministic DDRA.
    \item \textbf{Case~II (Issuer~B):} A credit card program where statement cycle based reward computation combined with auto redemption creates a timing window that produces an analogous reward integrity violation.
    \item \textbf{Case~III (Issuers~C, D, E, and F):} Credit card programs exhibiting a range of negative balance designs, from fully robust indefinite enforcement (Issuers~C, D, and E) to a partially robust design where instant reward availability combined with batched clawback creates a timing asymmetry (Issuer~F).
  \end{itemize}
  Our empirical scope reflects the accounts we personally hold and were not selected through systematic sampling. The six issuers span both debit and credit card products from major U.S. financial institutions. The small account scope reflects regulatory and identity verification barriers inherent to U.S. financial account creation, not a limitation of experimental design.

  \item \textbf{Formal model and threat framing.}
  We model reward engines as state machines over transactions and reward ledgers and define an adversary model that captures an honest but opportunistic consumer who uses only documented features (purchases, refunds, redemption) but strategically chooses timing.

  \item \textbf{Integrity invariants and vulnerability taxonomy.}
  We introduce two key security properties, \emph{Reward Integrity} and \emph{Refund Reward Consistency}, and identify several classes of reward logic flaws. We map these to CWE\nobreakdash-841 ~\cite{cwe841} and the recently published OWASP Business Logic Abuse Top~10~\cite{owaspbla2025}.

  \item \textbf{Defensive algorithms.}
  We propose implementation oriented pseudo algorithms with a semi formal correctness argument under stated assumptions. The algorithms link rewards to specific transactions, apply proportional clawback on full and partial refunds, support negative reward balances with automatic future offset, and reconcile statement cycle rewards against pending refunds.
\end{enumerate}

Our primary goal is not to expose or criticize specific providers but to elevate cashback reward logic to a first class security topic. We anonymize all providers and omit implementation details that would make identification trivial. We argue that reward engines should be engineered with explicit invariants and state transitions, much like payment authorization logic, and that security reviews should treat reward abuse attacks as seriously as more traditional attack classes.

The remainder of this paper is organized as follows. Section~\ref{sec:background} provides background on reward ecosystems, returns abuse, and business logic vulnerabilities. Section~\ref{sec:system-model} presents our system and threat model. Section~\ref{sec:invariants} formalizes integrity invariants and introduces the vulnerability taxonomy. Section~\ref{sec:case-studies} details our three case studies. Section~\ref{sec:defensive} presents the formal model and defensive algorithms. Section~\ref{sec:discussion} discusses practical considerations and economic impact. Section~\ref{sec:ethics} covers ethics, disclosure, and methodology. Section~\ref{sec:related} surveys related work. Section~\ref{sec:conclusion} concludes.

% ============================================================
% 2  BACKGROUND
% ============================================================
\section{Background}
\label{sec:background}

Modern payment ecosystems combine card networks, digital wallets, and merchant offers into layered reward structures that operate on top of core payment and refund rails. Consumers routinely earn cashback or loyalty points from their card issuer, and these reward artifacts behave like a soft currency whose lifecycle must track the economic lifecycle of the underlying transaction through authorization, settlement, partial or full refunds, chargebacks, and dispute resolution.

\subsection{Card Rewards and Cashback Ecosystems}
\label{sec:bg-rewards}

Card issuers and digital wallets typically implement reward engines that compute benefits as a function of transaction attributes such as merchant category, channel, and spend volume. Common structures include flat rate cashback (for instance, 1 to 2 percent on all purchases), rotating high reward categories (for instance, 5 percent on selected merchants up to a monthly cap), and travel points. These engines may treat refunds and chargebacks as separate transaction events, and must decide when to make rewards available for redemption and how to adjust rewards when the underlying spend is reversed.

The marketing and operations literature treats cashback primarily as a pricing and promotion instrument. Taleizadeh et al.\ analyze how cashback strategies affect sales under refund policies and customer credit, modeling the joint decisions of price, refund terms, and cashback intensity~\cite{taleizadeh2023cashback}. In subsequent work, the same group study retail pricing, cashback, and refund decisions in a supply chain with both online and direct channels, highlighting the strategic interactions between channel structure, cashback intensity, and refund policies~\cite{taleizadeh2024pricing}. Wu et al.\ examine a ``praise cashback'' strategy, where cashback is tied to positive online reviews, and quantify its implications for both consumers and online businesses~\cite{wu2023praise}. These papers model cashback as a rational incentive mechanism but generally assume compliant consumers and do not treat the reward engine as a security critical state machine that could be abused.

\subsection{Product Returns and Refund Abuse}
\label{sec:bg-returns}

Product returns and refund abuse have become a major operational concern. Product returns in the United States alone have been estimated in the hundreds of billions of dollars annually, with return rates for online fashion often exceeding 30 to 40 percent~\cite{frei2020returns}. Harris provides one of the earliest focused studies of fraudulent consumer returns, documenting how consumers exploit liberal return policies and identifying psychological and situational factors that increase such behavior~\cite{harris2010fraudulent}. Akturk et al.\ explicitly model return abuse, both opportunistic and fraudulent, and study how retailers can use technology enabled countermeasures such as customer profiling and product tracking to mitigate abuse~\cite{akturk2021managing}. Ketzenberg et al.\ use transaction level data to characterize customer return behaviors, segmenting customers by return propensity~\cite{ketzenberg2020returns}.

More recent work incorporates return policy abuse as a strategic behavior. Liu and Du focus on e commerce platforms' return policies under consumer abuse~\cite{liu2023returnpolicy}. Zhang et al.\ build a framework of fraudulent returns and mitigation strategies in multichannel retailing~\cite{zhang2023fraudulentreturns}. Merlano et al.\ study how consumers perceive anti fraud measures in omnichannel fashion retail~\cite{merlano2024fraudulentreturns}. Frei et al.\ explore how the COVID 19 pandemic affected returns management practices~\cite{frei2023covidreturns}. Collectively, this literature shows that return abuse is economically significant and that consumers can behave strategically. However, most of these works treat reward programs, if present at all, as exogenous context rather than as a detailed accounting mechanism whose state transitions interact with returns.

\subsection{Double Dip Rewards and Cashback Abuse}
\label{sec:bg-doubledip}

The combination of cashback incentives and liberal return policies creates natural opportunities for financial arbitrage. A user can earn a benefit at the moment of purchase, then later unwind the underlying economic exposure via a refund while attempting to retain the reward. If the reward engine does not treat refunds as first class events, this leads to a double dip pattern: the user briefly commits to a transaction, earns a reward, then exits the transaction while keeping the reward.

Promotion focused research shows that cashback and other incentives can themselves become targets of abuse. Vieira et al.\ investigate how cashback strategies generate financial benefits for retailers through the mediating role of consumer program loyalty~\cite{vieira2022cashback}. Jiang et al.\ explore the management of sales agents and product returns while implementing safeguards against fake orders~\cite{jiang2025fakeorder}. Aprisadianti and Dwiyanti\ develop an application to detect promotion abuse fraud utilizing a risk scoring methodology~\cite{aprisadianti2023promotion}. Li et al.\ introduce PromoGuardian, a graph based fraud detection system deployed at scale, which targets ``stocking up'' and cashback abuse patterns that are hard to distinguish from normal purchasing~\cite{li2025promoguardian}. Sun et al.\ show that concession abuse (repeated exploitation of customer service credits, refunds, and goodwill gestures) has been industrialized as ``Concession Abuse as a Service,'' with specialized communities sharing scripts and tactics~\cite{sun2021concession}.

\subsection{Business Logic Vulnerabilities}
\label{sec:bg-blv}

Application business logic vulnerabilities arise when an attacker can use a system exactly as designed at the interface level, but in a sequence or combination of actions that violate the underlying business rules. Unlike classic bugs such as SQL injection or buffer overflows, logic vulnerabilities involve no malformed input or protocol violation. Instead, they exploit gaps in how developers modeled the system's state machine: missing transitions, unchecked assumptions, or corner cases that product teams did not anticipate.

Standards bodies and security communities have recognized business logic issues for over a decade. MITRE's CWE\nobreakdash-841, Improper Enforcement of Behavioral Workflow, covers cases where a product supports multi step workflows but fails to ensure correct ordering of steps or correct state updates on transitions~\cite{cwe841}. In 2025, OWASP released the first Business Logic Abuse Top~10, which uses a Turing machine inspired model to categorize business logic vulnerabilities into ten classes including lifecycle and orphaned transition flaws, sequential state bypass, and transition validation flaws~\cite{owaspbla2025}. This release signals growing recognition that business logic abuse constitutes a distinct and expanding threat category.

Research has proposed various techniques to discover such flaws. Felmetsger et al.\ proposed one of the first automated approaches to detect logic vulnerabilities by inferring expected workflows from dynamic execution~\cite{felmetsger2010logic}. Pellegrino and Balzarotti demonstrated a black box technique that learns navigation graphs from network traces~\cite{pellegrino2014blackbox}. Chen et al.\ examined logic vulnerabilities in payment syndication services, showing how inconsistencies between documentation and implementation can produce exploitable behavior~\cite{chen2019devils}. Wang et al.\ analyzed Cashier as a Service web stores and showed that integration errors can allow attackers to manipulate prices or obtain goods without paying~\cite{wang2011shopfree}. Ghorbansadeh and Shahriari proposed detecting application logic vulnerabilities by finding incompatibilities between design and implementation~\cite{ghorbansadeh2020detecting}.

Our work differs from these prior strands in two key respects. First, we do not focus on collusive schemes or large scale fake orders: our adversary model considers single, honest but opportunistic consumers who exploit legitimate purchases and refunds. Second, rather than building another detection system, we focus on the internal reward engine logic itself and show that small design flaws are sufficient to create reward abuse attack opportunities that violate basic integrity invariants.

% ============================================================
% 3  SYSTEM AND THREAT MODEL
% ============================================================
\section{System and Threat Model}
\label{sec:system-model}

We now describe the environment in which reward engines operate and the type of adversary we consider. Our goal is to define a minimal model that is rich enough to express the case studies in Section~\ref{sec:case-studies} and the invariants in Section~\ref{sec:invariants}.

\subsection{System Entities}

We model a generic reward ecosystem with the following entities:

\begin{itemize}[leftmargin=*]
  \item \textbf{User} $U$: a legitimate consumer who participates in one or more card or loyalty programs. A single user may hold multiple cards and maintain separate reward ledgers with different issuers.

  \item \textbf{Issuer} $I$: a bank, fintech company, or wallet provider that issues cards or maintains a reward ledger for users. Issuers define earning rules, redemption channels, and caps, and are responsible for implementing the reward engine logic we study.

  \item \textbf{Merchant} $M$: a retailer or service provider that accepts payments. A merchant can appear as a physical point of sale, an e commerce site, or a marketplace platform.

  \item \textbf{Payment Network} $N$: card networks and processors that handle authorization, clearing, and settlement. We treat $N$ as honest and reliable; it determines which transactions and refunds are visible to each issuer.
\end{itemize}

\begin{figure}[t]
\centering
\begin{tikzpicture}[
  node distance=1.4cm and 2.2cm,
  every node/.style={font=\small},
  entity/.style={draw, rounded corners=6pt, minimum width=1.8cm, minimum height=0.85cm, thick, font=\small\bfseries},
  arrow/.style={-{Stealth[length=2.5mm,width=2mm]}, thick, color=darkgray},
  lbl/.style={font=\scriptsize, midway, fill=white, inner sep=1.5pt, text=darkgray}
]
  \node[entity, fill=softblue, draw=accentblue] (U) {User $U$};
  \node[entity, fill=softgreen, draw=accentgreen, right=3cm of U] (M) {Merchant $M$};
  \node[entity, fill=softteal, draw=accentteal, below=1.8cm of U] (I) {Issuer $I$};
  \node[entity, fill=softamber, draw=accentamber, below=1.8cm of M] (N) {Network $N$};

  \draw[arrow, accentblue] (U) -- node[lbl, above] {purchase / return} (M);
  \draw[arrow, accentgreen] (M) -- node[lbl, right] {settle / refund} (N);
  \draw[arrow, accentamber] (N) -- node[lbl, below] {txn events} (I);
  \draw[arrow, accentteal] (I) -- node[lbl, right] {rewards} (U);
  \draw[arrow, dashed, accentred] (I) to[bend left=20] node[lbl, left, pos=0.35, text=accentred] {\emph{adjust?}} (U);
\end{tikzpicture}
\caption{Reward ecosystem entities and event flows. Solid arrows represent primary transaction and reward flows. The dashed red arrow represents the reward adjustment path that is absent in Case~I (V1) and absent for cross cycle refunds in Case~II (V2).}
\label{fig:ecosystem}
\end{figure}
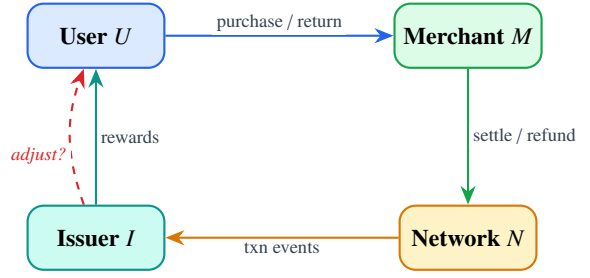

Figure~\ref{fig:ecosystem} illustrates the relationships among these entities and the flow of transaction, refund, and reward events.

Issuers maintain a reward ledger $B[U]$ for each user, representing cashback or points that may be convertible to cash, statement credit, or vouchers. We refer to the logic that maps observed events (purchases, refunds) to updates on this ledger as the \emph{reward engine}.

\subsection{Reward Engine Abstraction}
\label{sec:reward-abstraction}

We abstract rewards as simple functions over transactions and events, intentionally omitting implementation details such as database schemas or APIs.

Each \textbf{transaction} $t$ is a tuple:
\[
t = (t.\mathit{id},\; t.u,\; t.m,\; t.a,\; t.c,\; t.n,\; t.\mathit{status}),
\]
where $t.u$ is the user, $t.m$ the merchant, $t.a$ the amount, $t.c$ a category label (e.g., \texttt{GROCERY}, \texttt{GAS}), $t.n$ the billing period in which the transaction settles, and
\begin{multline*}
t.\mathit{status} \in \{\tsc{PENDING}, \tsc{SETTLED}, \tsc{PART\_REF},\\
\tsc{REFUNDED}, \tsc{CHARGEBACK}\}.
\end{multline*}
We assume that once a transaction reaches \tsc{SETTLED}, its amount and category do not change; refunds and chargebacks are modeled as separate events attached to $t$.

For each transaction, we maintain a scalar $R[t.\mathit{id}]$ that records the total reward value granted for that transaction so far. In defensive implementations, $R[t.\mathit{id}]$ may decrease over time as refunds are processed.

A \textbf{reward engine} implements three core operations:

\begin{itemize}[leftmargin=*]
  \item $\mathsf{RewardRate}(c)$: the reward percentage for category $c$. This function implicitly captures caps and eligibility rules.

  \item $\mathsf{OnSettled}(t)$: invoked when a transaction becomes settled and reward eligible. This is where rewards are initially computed and credited.

  \item $\mathsf{OnRefund}(t,x)$: invoked when an amount $x$ of transaction $t$ is refunded or charged back. This is where previously granted rewards should be adjusted.
\end{itemize}

The reward balance $B[U]$ is a numeric ledger that accumulates contributions across all settled transactions and redemptions. We explicitly allow $B[U]$ to become negative in some designs, which is essential for safely handling late refunds after rewards have been redeemed.

\subsection{Event Timelines}

Transactions and promotions unfold over time. For a single purchase as seen by an issuer, a typical timeline proceeds through five stages:

\begin{enumerate}[leftmargin=*]
  \item At time $t_{\text{auth}}$, the user presents a card or wallet; an authorization event is processed by $N$.
  \item At time $t_{\text{settle}}$, a settlement event with final amount $t.a$ and merchant $t.m$ is delivered to $I$.
  \item At time $t_{\text{reward}}$, the issuer's reward engine invokes $\mathsf{OnSettled}(t)$, updating $B[U]$ and $R[t.\mathit{id}]$. In some designs, rewards become immediately redeemable; in others, they remain pending until a statement cycle ends.
  \item At some later time $t_{\text{refund}}$, the user may return goods; $M$ issues a refund that flows through $N$ to $I$.
  \item At time $t_{\text{adjust}}$, if implemented, the issuer invokes $\mathsf{OnRefund}(t,x)$, adjusting $B[U]$ and $R[t.\mathit{id}]$ accordingly.
\end{enumerate}

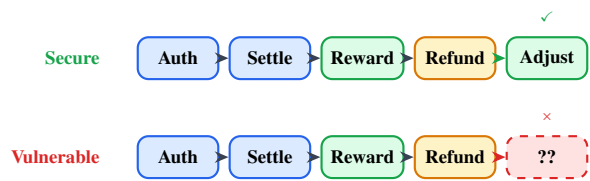
\begin{figure}[t]
\centering
\resizebox{\columnwidth}{!}{%
\begin{tikzpicture}[
  node distance=0.12cm,
  every node/.style={font=\scriptsize},
  phase/.style={draw, rounded corners=4pt, minimum height=0.55cm, minimum width=1.05cm, thick, text centered, font=\scriptsize\bfseries},
  arr/.style={-{Stealth[length=1.8mm, width=1.5mm]}, thick, color=darkgray}
]
  % Secure timeline
  \node[font=\scriptsize\bfseries, anchor=east, text=accentgreen] at (0.2, 1.2) {Secure};
  \node[phase, fill=softblue, draw=accentblue] (s1) at (1.1, 1.2) {Auth};
  \node[phase, fill=softblue, draw=accentblue, right=0.12cm of s1] (s2) {Settle};
  \node[phase, fill=softgreen, draw=accentgreen, right=0.12cm of s2] (s3) {Reward};
  \node[phase, fill=softamber, draw=accentamber, right=0.12cm of s3] (s4) {Refund};
  \node[phase, fill=softgreen, draw=accentgreen, right=0.12cm of s4] (s5) {Adjust};
  \draw[arr] (s1) -- (s2);
  \draw[arr] (s2) -- (s3);
  \draw[arr] (s3) -- (s4);
  \draw[arr, color=accentgreen] (s4) -- (s5);
  \node[font=\tiny, text=accentgreen, above=0.08cm of s5] {\checkmark};

  % Vulnerable timeline
  \node[font=\scriptsize\bfseries, anchor=east, text=accentred] at (0.2, -0.1) {Vulnerable};
  \node[phase, fill=softblue, draw=accentblue] (v1) at (1.1, -0.1) {Auth};
  \node[phase, fill=softblue, draw=accentblue, right=0.12cm of v1] (v2) {Settle};
  \node[phase, fill=softgreen, draw=accentgreen, right=0.12cm of v2] (v3) {Reward};
  \node[phase, fill=softamber, draw=accentamber, right=0.12cm of v3] (v4) {Refund};
  \node[phase, fill=softred, draw=accentred, right=0.12cm of v4, dashed] (v5) {\textbf{??}};
  \draw[arr] (v1) -- (v2);
  \draw[arr] (v2) -- (v3);
  \draw[arr] (v3) -- (v4);
  \draw[arr, dashed, color=accentred] (v4) -- (v5);
  \node[font=\tiny, text=accentred, above=0.08cm of v5] {$\times$};
\end{tikzpicture}
}%
\caption{Transaction reward lifecycle for the V1 vulnerability class (Case~I). In secure systems (top), refund events trigger proportional reward adjustment. In vulnerable systems (bottom), step~5 is absent entirely. In V2 (Case~II), step~5 exists within the billing cycle but not across cycle boundaries; see Figure~\ref{fig:case2-timing}. The intermediate tier (V3, Issuer~F) where step~5 fires but with timing asymmetry is detailed in Section~\ref{sec:case3b}.}
\label{fig:timeline}
\end{figure}

Figure~\ref{fig:timeline} contrasts the transaction reward lifecycle under a vulnerable system versus a secure system. The critical distinction is the presence or absence of step~5.

\subsection{Adversary Model}
\label{sec:adversary}

We consider an honest but opportunistic \emph{reward abuse attacker} whose capabilities mirror those of a typical consumer, with a bias toward systematic exploitation of reward programs.

The adversary can initiate legitimate purchases and refunds via normal merchant channels, including full and partial returns where the merchant's policy allows. The adversary can choose among different payment instruments such as different cards or wallets. Critically, the adversary can time actions strategically: redeeming rewards as soon as they become available and then waiting to request refunds, or scheduling returns to fall in a different billing cycle than the original purchase. The adversary may hold multiple legitimate accounts with different providers, subject to normal onboarding checks. Our model assumes at most one account per issuer; multiple same issuer accounts scale the financial impact linearly but do not change the fundamental vulnerability class or the correctness argument for the defensive algorithms.

The adversary does \emph{not} compromise backend systems at issuers or merchants, forge or tamper with payment network messages, use stolen cards or identities, make false claims about non receipt or product condition, or exploit low level implementation bugs such as SQL injection or memory corruption.

The adversary's goal is to increase net financial gain from rewards beyond what program rules intend, specifically to obtain net rewards inconsistent with net spend, while minimizing detection and account closure risk.

We note that prior work documents industrialized, scripted exploitation of concession and refund workflows~\cite{sun2021concession}, suggesting that automated adversaries operating at higher velocity are realistic. Our adversary model focuses on the single user manual tier because it represents the minimum capability needed to exploit the vulnerabilities we describe. Importantly, the defensive algorithms we propose in Section~\ref{sec:defensive} operate as stateless, per event checks (per transaction reward tracking, proportional clawback on each refund event, negative balance enforcement on each redemption request). These mechanisms function identically regardless of whether the refund events arrive from a single manual user or from an automated script generating hundreds of refund requests per hour.

\subsection{Security Properties}

We focus on two main security properties. \textbf{Reward Integrity} requires that the total economic value a user derives from rewards should never exceed what they are entitled to based on their net legitimate spending and published program rules. \textbf{Refund Reward Consistency} requires that when purchases are reversed via refunds or chargebacks, the reward state (balances, per transaction records) should be updated within a bounded time to remain consistent with the new economic reality. These are formalized in Section~\ref{sec:invariants}. We do not aim to model every form of fraud; instead, we focus on design flaws in reward engines that even a single honest but opportunistic user can exploit. We treat higher level fraud analytics and anomaly detection systems as orthogonal: they may detect or throttle some abuse patterns, but they cannot restore correctness if the core reward logic violates basic invariants.

\subsection{STRIDE Perspective}
\label{sec:stride}

From a STRIDE perspective, the vulnerabilities we study primarily affect the \emph{integrity} of reward ledgers (Tampering) and, secondarily, the ability of issuers to justify reward adjustments with clear audit trails (Repudiation). Spoofing of user identities, information disclosure, denial of service, and elevation of privilege are outside our adversary model. We do not mark Elevation of Privilege because the adversary remains within their authorized user role throughout; the flaw is a ledger integrity violation (Tampering) rather than a privilege boundary crossing in the STRIDE sense. This aligns with CWE\nobreakdash-841~\cite{cwe841}, as well as the OWASP Business Logic Abuse Top~10 classes covering lifecycle and orphaned transition flaws and transition validation flaws~\cite{owaspbla2025}. For Issuer~F (Case~III-B), the ``partial'' annotations in Table~\ref{tab:stride} reflect that the reward ledger is eventually updated (the negative entry is created), mitigating the Tampering dimension under normal continued usage, but the timing asymmetry between instant redemption and batched clawback leaves a window during which the ledger state does not reflect the true economic position, and the issuer may lack a cost effective mechanism to repudiate the already issued credit if the user ceases card usage.

Table~\ref{tab:stride} summarizes how our case studies map to STRIDE categories.

\begin{table}[t]
\scriptsize
\centering
\caption{STRIDE mapping of case studies. Checkmarks indicate primary categories impacted under our adversary model. ``--'' denotes categories outside the adversary model scope.}
\label{tab:stride}
\resizebox{\columnwidth}{!}{%
\begin{tabular}{@{}lcccccc@{}}
\toprule
\textbf{Case} & \textbf{S} & \textbf{T} & \textbf{R} & \textbf{I} & \textbf{D} & \textbf{E} \\
\midrule
I (Issuer A)       &  -- & \checkmark & \checkmark & --  &  -- &  -- \\
II (Issuer B)        &  -- & \checkmark & \checkmark & --  & --  &  -- \\
III-A (C,D,E)  & --  & mitigates  &    --        &  -- & --  &  -- \\
III-B (Issuer F)       &  -- & partial    & partial    &  -- &  -- & --  \\
\bottomrule
\end{tabular}%
}
% {\footnotesize S=Spoofing, T=Tampering, R=Repudiation,
% I=Information Disclosure, D=Denial of Service,
% E=Elevation of Privilege.}
\end{table}

% ============================================================
% 4  INVARIANTS AND VULNERABILITY CLASSES
% ============================================================
\section{Reward Logic Invariants and Vulnerability Classes}
\label{sec:invariants}

We now formalize what it means for a reward system to behave correctly with respect to user spend and refunds, and classify the main ways in which real world systems violate these properties.

\subsection{Formal Definitions}
\label{sec:integrity}

Let $\mathcal{T}(U)$ denote the set of all transactions involving user $U$ as observed by a particular reward engine.

\paragraph{Net Spend.}
For each $t \in \mathcal{T}(U)$, let $x_t$ denote the total amount refunded or charged back for that transaction (possibly $0$). We define the \emph{net spend} of user $U$ as:
\begin{equation}
  \label{eq:net-spend}
  \mathsf{NetSpend}(U)
  = \sum_{t \in \mathcal{T}_{\text{settled}}(U)} (t.a - x_t),
\end{equation}
where $\mathcal{T}_{\text{settled}}(U)$ denotes all transactions that have reached at least \tsc{SETTLED} status (including \tsc{PART\_REF}, \tsc{REFUNDED}, and \tsc{CHARGEBACK}, all of which were settled before being reversed), and $x_t = \mathsf{TotalRefunded}[t.\mathit{id}]$ is the cumulative refund amount as of the evaluation time. Equation~\eqref{eq:net-spend} is a snapshot quantity: for a \tsc{PART\_REF} transaction, $x_t$ reflects the refunds processed so far and may increase as additional partial refunds arrive. For a fully refunded transaction, $x_t = t.a$ and the contribution is zero.

\paragraph{Net Reward.}
Let $B[U]$ be the current reward balance and let $R_{\text{red}}(U)$ be the total value of rewards already redeemed. We define:
\begin{equation}
  \label{eq:net-reward}
  \mathsf{NetReward}(U) = B[U] + R_{\text{red}}(U).
\end{equation}

\paragraph{Reward Integrity.}
A reward system satisfies \emph{Reward Integrity} if, for all users $U$:
\begin{equation}
  \label{eq:integrity}
  \mathsf{NetReward}(U)
  \;\le\;
  f\bigl(\mathsf{NetSpend}(U),\, \mathsf{PromoRules}\bigr),
\end{equation}
where $f(\cdot)$ encodes the program's published reward rates, per category caps, and promotions. For a simple 1 percent cashback program with a single category, $f(s,\cdot) = 0.01 \cdot \max(s, 0)$. For multi category programs with per category rates and caps, $f$ is understood as a sum over categories $c$, each applying its own rate and cap to the net spend in that category; we omit the per category decomposition for brevity since our algorithms operate per transaction with category labels $t.c$.

Case~I is precisely a violation: refunds drive $\mathsf{NetSpend}(U)$ toward zero, but $\mathsf{NetReward}(U)$ remains positive.

\paragraph{Refund Reward Consistency.}
A reward system satisfies \emph{Refund Reward Consistency} if, for any transaction $t$ of user $U$, whenever a nonzero refund amount $x_t > 0$ is processed, there exists a bounded time $\Delta$ such that after at most $\Delta$ time:
\begin{enumerate}[leftmargin=*]
  \item the state of $t$ reflects the refund ($t.\mathit{status} \in \{\tsc{REFUNDED}, \tsc{CHARGEBACK}\}$), and
  \item the reward state has been updated so that condition~\eqref{eq:integrity} still holds.
\end{enumerate}

This allows for asynchronous or batched processing but rules out systems where rewards are never adjusted.

\subsection{Additional Invariants}

Implementations may enforce further properties:

\begin{itemize}[leftmargin=*]
  \item \textbf{Per Transaction Bound.} For each transaction $t$, the total reward should never exceed a known function of $t.a$ and $t.c$.

  \item \textbf{Monotonicity with Net Spend.} If a user's net spend does not increase over a time window, their net reward should not increase either.
\end{itemize}

\subsection{Vulnerability Taxonomy}
\label{sec:vuln-classes}

From our modeling and case studies, we identify three primary classes of reward logic flaws relevant to refund based reward abuse:

\begin{enumerate}[leftmargin=*]
  \item \textbf{V1: Refund Insensitive Rewards.}
  Rewards are granted and become redeemable, but are never adjusted when the underlying transaction is refunded. The reward engine has no refund event handler or refund aware code path. The reward ledger behaves as if only positive settlement events exist. This directly violates both Reward Integrity and Refund Reward Consistency (Case~I).

  \item \textbf{V2: Temporally Scoped Adjustment.}
  The reward engine includes a refund aware computation mechanism, but its scope is bounded to a specific temporal window such as the current billing cycle. Refund events that fall outside this window are not processed by the adjustment mechanism, creating a permanent gap for out of scope refunds. If rewards are made available for redemption before the scope window closes, users can redeem and then initiate refunds that land outside the adjustment window. This differs from V1 in that a within scope adjustment mechanism exists; it differs from V3 in that out of scope refunds produce zero clawback rather than a flawed clawback. V2 is a temporal scoping flaw: the security relevant state (net spend) changes after the reward computation event but before the adjustment mechanism's scope window can intercept it. This is structurally distinct from TOCTOU, which involves concurrency; V2 arises from a deterministic architectural mismatch between billing cycle boundaries and merchant return windows. This maps to CWE\nobreakdash-841 (Case~II).

  \item \textbf{V3: Inconsistent Negative Balance Handling.}
  Refunds can push balances negative, but the system does not consistently offset future rewards against the negative balance. V3 manifests in two forms: (a) no negative balance support at all, where the system simply ignores the reward debt; and (b) a negative balance is correctly created but a timing asymmetry between instant reward availability and batched clawback allows the user to extract value before the adjustment occurs, and recovery depends entirely on continued card usage that the user can unilaterally withhold. Form~(a) is included for taxonomic completeness; we did not observe a pure V3(a) instance in our sample, though it is architecturally plausible for systems that cap negative balances at zero. V3(a) is structurally distinct from V1: in V3(a), an $\mathsf{OnRefund}$ handler exists and fires correctly, but the balance update discards the computed debt by flooring at zero rather than permitting a negative balance. The observable outcome after redemption is identical to V1, but the root cause and required fix differ: V1 requires building a handler; V3(a) requires only removing the zero floor guard. Case~III illustrates both robust and partially robust approaches.
\end{enumerate}

Table~\ref{tab:vuln-taxonomy} maps each class to the invariants it violates, relevant standards, and our case studies.

\begin{table}[t]
\small
\centering
\caption{Vulnerability taxonomy with mappings to invariants, CWE/OWASP categories, and case studies. RI = Reward Integrity, RRC = Refund Reward Consistency.}
\label{tab:vuln-taxonomy}
\resizebox{\columnwidth}{!}{%
\begin{tabular}{@{}lllll@{}}
\toprule
\textbf{Class} & \textbf{Invariant} & \textbf{CWE} & \textbf{OWASP BLA Class} & \textbf{Case} \\
\midrule
V1: Refund Insensitive & RI, RRC & 841 & Class 1 & I \\
V2: Temporally Scoped & RRC & 841 & Class 7 & II \\
V3: Neg.\ Balance & RI & 841 & Class 4 & III-B \\
\bottomrule
\end{tabular}%
}
\end{table}

Violations of these invariants correspond to CWE\nobreakdash-841 (improper enforcement of behavioral workflow) under parent category CWE-840 (Business Logic Errors), as the system fails to enforce the required sequence of: purchase $\rightarrow$ reward $\rightarrow$ refund $\rightarrow$ reward adjustment ~\cite{cwe841}. In the OWASP
Business Logic Abuse Top~10, V1 aligns with Class~1
(Lifecycle \& Orphaned Transition Flaws), where the refund
to reward adjustment transition is entirely absent; V2
aligns with Class~7 (Transition Validation Flaws), where a
within-scope adjustment mechanism exists but its temporal
coverage is insufficient to handle cross-cycle refunds;
V3(b) aligns with Class~4 (Sequential State Bypass), where
the redemption step proceeds before the required clawback
settlement has completed~\cite{owaspbla2025}.

% ============================================================
% 5  CASE STUDIES
% ============================================================
\section{Case Studies}
\label{sec:case-studies}

We now present three case studies that instantiate the vulnerability classes above. In each case we used small, legitimate transactions on accounts we personally control and followed normal user workflows. All providers are anonymized.

\subsection{Case I: Refund Insensitive Debit Cashback (Issuer A)}
\label{sec:case1}

\paragraph{Context}
Issuer~A operates a debit based cashback program in which users earn 5 percent cashback on purchases in selected categories, subject to a monthly cap equivalent to approximately \$50 per month. Users can enable a specific category each month (for example, groceries, restaurants, or general retail). Rewards are credited shortly after the transaction posts and become immediately redeemable as a cash equivalent balance that can be transferred to any linked account.

\paragraph{Observed Behavior}
Using a personal account, we performed low value test purchases at merchants within the 5 percent category. We waited until the cashback appeared and became available for redemption, which typically occurred within minutes to hours of settlement. We then redeemed the earned rewards by converting them to cash equivalent balance and transferred the funds to another account. Finally, we initiated legitimate refunds through standard merchant return processes, including both full and partial returns where the merchant's policy allowed.

For all tested merchants, when the refund posted to the account, the principal amount was correctly returned to the debit balance. However, no adjustment was made to the reward ledger: previously granted rewards were neither debited nor marked as pending reversal. Future rewards accrued normally and were completely unaffected by the refunded transactions. After a full refund, the user's net spend on those purchases returned to approximately zero, while net rewards remained positive. This constitutes a direct violation of Reward Integrity.

\begin{figure}[t]
\centering
\resizebox{\columnwidth}{!}{%
\begin{tikzpicture}[
  node distance=0.6cm,
  every node/.style={font=\scriptsize},
  phase/.style={draw, rounded corners=5pt, minimum height=0.7cm, minimum width=2.8cm, thick, text centered, font=\scriptsize\bfseries},
  arr/.style={-{Stealth[length=2mm, width=1.8mm]}, very thick},
  note/.style={font=\tiny, text=darkgray, align=center}
]
  \node[phase, fill=softblue, draw=accentblue] (p1) {Phase 1: Earn Rewards};
  \node[phase, fill=softgreen, draw=accentgreen, below=0.5cm of p1] (p2) {Phase 2: Extract Cash};
  \node[phase, fill=softamber, draw=accentamber, below=0.5cm of p2] (p3) {Phase 3: Return \& Refund};
  \node[phase, fill=softred, draw=accentred, below=0.5cm of p3] (p4) {Phase 4: Retain Rewards};

  \draw[arr, accentblue] (p1) -- (p2);
  \draw[arr, accentgreen] (p2) -- (p3);
  \draw[arr, accentamber] (p3) -- (p4);

  \node[note, right=0.25cm of p1] {Buy in 5\% category};
  \node[note, right=0.25cm of p2] {Redeem \& transfer out};
  \node[note, right=0.25cm of p3] {Merchant issues refund};
  \node[note, right=0.25cm of p4, text=accentred] {Net spend $\approx$ \$0\\Rewards $>$ \$0};

  \draw[arr, dashed, accentred, thick] (p4.west) to[out=180, in=180] node[note, left, text=accentred] {repeat\\monthly} (p1.west);
\end{tikzpicture}
}%
\caption{The four phase double dip cashback reward abuse attack (DDRA) workflow for Case~I. The dashed arrow indicates that this cycle can be repeated each billing period up to the reward cap, yielding deterministic profit with zero net spend.}
\label{fig:ddpa-flow}
\end{figure}
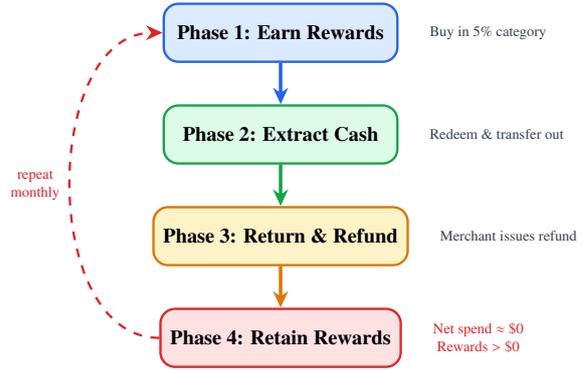

\paragraph{Four Phase Attack Workflow}
The attack follows a deterministic four phase pattern, illustrated in Figure~\ref{fig:ddpa-flow}. In the first phase (\emph{Earn Rewards}), the attacker purchases items in the bonus category and waits for the cashback to post, typically within minutes to hours of settlement. In the second phase (\emph{Extract Cash}), the attacker redeems the accrued cashback into a spendable balance and transfers it to an external account, removing the value from the reward ledger's reach. In the third phase (\emph{Return \& Refund}), the attacker returns the purchased items through the merchant's standard return process and receives a refund of the principal amount to the card. In the fourth phase (\emph{Retain Rewards}), the attacker observes that the reward ledger has not been adjusted: the net spend has returned to approximately zero, but the redeemed cashback remains in the external account. Because the reward engine does not process refund events, the cycle can be repeated each billing period up to the monthly reward cap.

\paragraph{Classification}
This vulnerability falls under CWE\nobreakdash-841 (Improper Enforcement of Behavioral Workflow) within the CWE\nobreakdash-840 Business Logic Errors category. The expected workflow ``Purchase $\rightarrow$ RewardGrant $\rightarrow$ Refund $\rightarrow$ RewardAdjust'' is not enforced: the RewardAdjust step is entirely absent. In the OWASP Business Logic Abuse Top~10, this maps to Class~1 (Lifecycle \& Orphaned Transition Flaws), as the refund to reward adjustment transition is missing entirely~\cite{owaspbla2025}.

\paragraph{Impact Estimation}
If the monthly cap is \$50 and the program covers a population of $|U|$ active users, a fraction $p$ of whom learn to systematically exploit this behavior, the direct annual reward leakage is approximately $p \cdot |U| \cdot \$600$. As an illustrative example, at $p = 0.01$ and $|U| = 1{,}000{,}000$, this yields roughly \$6~million per year; however, neither $p$ nor $|U|$ is empirically estimated from our data, and this figure should be interpreted as a sensitivity bound rather than a prediction of actual losses.

\begin{takeawaybox}
\textbf{Case I Takeaway.}
Refund insensitive cashback accounting allows users to redeem rewards on purchases that are later fully refunded, creating a deterministic double dip attack using legitimate accounts and standard refund workflows.
\end{takeawaybox}

\subsection{Case II: Statement Cycle Timing Exploit (Issuer B)}
\label{sec:case2}

\paragraph{Context}
Issuer~B operates a credit card program with a statement cycle based reward computation model. At the close of each billing cycle, the issuer evaluates all transactions from that period, computes rewards based on category spend, and credits the corresponding cashback. The program offers an auto redemption feature that automatically converts accrued rewards into statement credits when the billing cycle closes. Many merchants that appear in the rewarded categories allow product returns within 30 calendar days of purchase.

\paragraph{Observed Behavior}
Using a personal account with the auto redemption feature enabled, we observed the following sequence across two billing periods. During billing period~$n$, we made purchases in a bonus category. At the close of period~$n$, the issuer computed rewards based on all transactions in that period and auto redeemed them as statement credits. During billing period~$n\!+\!1$, but still within the merchant's 30 day return window, we returned the purchased items and received refunds. The issuer processed the refund as a credit in period~$n\!+\!1$ but did not claw back the rewards that had already been computed and auto redeemed in period~$n$.

The timing gap between reward computation (end of period~$n$) and the return window (extending into period~$n\!+\!1$) creates an exploitable window. Once auto redemption has occurred, the rewards from period~$n$ transactions are treated as finalized and cannot be reversed even when the underlying purchases are subsequently returned.

\begin{figure}[t]
\centering
\begin{tikzpicture}[
  every node/.style={font=\scriptsize},
  timeline/.style={very thick, color=darkgray},
  tick/.style={thick, color=darkgray},
  event/.style={font=\scriptsize, text=darkgray},
  period/.style={draw, fill=softblue, rounded corners=3pt, minimum height=0.45cm, thick, draw=accentblue, font=\scriptsize\bfseries},
  window/.style={draw, fill=softamber, rounded corners=3pt, minimum height=0.35cm, thick, draw=accentamber, font=\tiny},
]
  % Timeline
  \draw[timeline] (0,0) -- (7.5,0);

  % Period markers
  \draw[tick] (0.5, 0.12) -- (0.5, -0.12) node[below, font=\tiny] {Stmt Open};
  \draw[tick] (3.5, 0.12) -- (3.5, -0.12) node[below, font=\tiny] {Stmt Close};
  \draw[tick] (6.5, 0.12) -- (6.5, -0.12) node[below, font=\tiny] {Period $n\!+\!1$ Close};

  % Periods
  \node[period] at (2.15, 0.7) {Period $n$};
  \node[period] at (5.8, 0.7) {Period $n\!+\!1$};

  % Events
  \node[font=\tiny, text=accentblue, fill=softblue, rounded corners=2pt, inner sep=2pt] (buy) at (1.5, 1.4) {Purchase};
  \draw[-{Stealth[length=1.5mm]}, accentblue, thick] (buy) -- (1.5, 0.15);

  \node[font=\tiny, text=accentgreen, fill=softgreen, rounded corners=2pt, inner sep=2pt] (rew) at (3.5, 1.4) {Reward + Auto Redeem};
  \draw[-{Stealth[length=1.5mm]}, accentgreen, thick] (rew) -- (3.5, 0.15);

  \node[font=\tiny, text=accentred, fill=softred, rounded corners=2pt, inner sep=2pt] (ret) at (5.0, 1.4) {Return \& Refund};
  \draw[-{Stealth[length=1.5mm]}, accentred, thick] (ret) -- (5.0, 0.15);

  % Return window
  \draw[{Stealth[length=1.2mm]}-{Stealth[length=1.2mm]}, accentamber, thick] (1.5, -0.55) -- (5.05, -0.55);
  \node[font=\tiny, text=accentamber] at (3.5, -0.8) {30 day return window};

  % Annotation
  \node[font=\tiny, text=accentred, align=center] at (6.05, -.89) {No clawback:\\rewards already finalized};
\end{tikzpicture}
\caption{Statement cycle timing vulnerability in Case~II. Rewards are computed and auto redeemed at the close of period~$n$. The merchant's 30 day return window extends into period~$n\!+\!1$, and the refund does not trigger any reward reversal.}
\label{fig:case2-timing}
\end{figure}
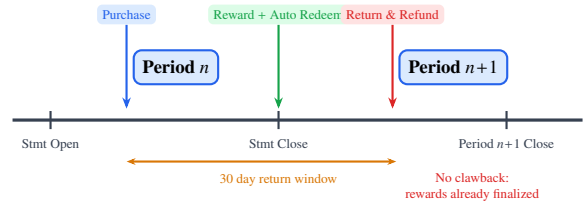

Figure~\ref{fig:case2-timing} illustrates this timing vulnerability across billing periods.

\paragraph{Classification}
This is an instance of V2 (Temporally Scoped Adjustment). Issuer~B's reward engine does compute rewards based on within period transaction activity, and same period refunds would reduce the reward eligible base at statement close. However, the system treats rewards as permanently finalized at each statement close and has no mechanism to retrospectively adjust rewards from a prior period when a cross cycle refund arrives. The vulnerability arises because the merchant's return window extends beyond the billing cycle boundary, placing the refund outside the reward engine's adjustment scope. This differs from V1 (Case~I), where the reward engine has no refund awareness at all, and from V3 (Case~III-B), where the adjustment mechanism fires across cycles but balance recovery is flawed.

\paragraph{Impact}
The potential leakage per exploitable period depends on the program's cap structure and may differ from Case~I. Because the reward computation is tied to the billing cycle rather than to individual transactions, the system lacks the per transaction linkage needed to perform proportional clawback after statement close.

\begin{takeawaybox}
\textbf{Case II Takeaway.}
Statement cycle based reward computation combined with auto redemption can create timing windows where rewards are finalized before the merchant return window closes, producing a reward integrity violation through a different mechanism than Case~I.
\end{takeawaybox}

\subsection{Case III: Negative Balance Designs (Issuers C, D, E, and F)}
\label{sec:case3}

To contrast with Cases I and II, we examined credit card reward programs from four additional issuers. Our observations reveal a spectrum of negative balance handling, ranging from fully robust indefinite enforcement to a partially robust design with a bounded recovery window.

\subsubsection{Case III-A: Robust Indefinite Enforcement (Issuers C, D, and E)}
\label{sec:case3a}

\paragraph{Context}
Issuers C, D, and E each operate credit card reward programs whose public documentation and observed behavior indicate robust refund handling. All three issuers state that rewards are computed on net purchases and may be reversed on returns, even resulting in negative reward balances.

\paragraph{Observed Behavior}
Using personal accounts and small transactions, we observed the core negative balance enforcement pattern at all three issuers. A purchase in a rewarded category posts and yields a positive reward increment. If the user redeems that reward (for example, as a statement credit), the reward balance reflects the redemption. When a refund later posts for the original transaction, the issuer debits the corresponding rewards, sometimes resulting in a negative balance if the user had already redeemed. The negative balance persists indefinitely: subsequent purchases first reduce the negative balance, and only after returning to a non negative balance does the user accumulate redeemable rewards again.

While all three issuers enforce the same indefinite negative balance netting outcome, they differ in their reward timing architecture. Issuer~C posts reward points shortly after transaction settlement and applies clawback immediately upon refund processing. Issuers~D and E compute and post reward points at statement cycle close, and apply refund adjustments at the same cadence. Despite this architectural difference, the security relevant behavior is identical: in all three cases, the negative balance is never forgiven or time bounded, and future earnings are automatically offset against any outstanding debt regardless of how many billing cycles elapse. This pattern directly enforces Reward Integrity.

\begin{takeawaybox}
\textbf{Case III-A Takeaway.}
Issuers that link rewards to individual transactions, apply proportional clawback on every refund, and enforce negative balances indefinitely against future earnings provide robust protection against refund based reward abuse. Three production issuers independently deploy this pattern with different reward timing architectures (instant and statement cycle), demonstrating its viability across both designs.
\end{takeawaybox}

\subsubsection{Case III-B: Timing Asymmetry in Reward Clawback (Issuer F)}
\label{sec:case3b}

\paragraph{Context}
Issuer~F operates a credit card reward program in which reward points become available for redemption as soon as the underlying transaction settles (typically within three business days of purchase). However, reward adjustments for refunds are not processed until statement close. This architectural choice creates a timing asymmetry: the user can access and redeem reward value in near real time, but the system does not reconcile refund based clawbacks until the end of the billing cycle.

\paragraph{Observed Behavior}
Using a personal account, we confirmed this asymmetry through a controlled sequence. We made a purchase in a rewarded category and observed that the reward points appeared in the redeemable balance within days of settlement, well before the current billing cycle closed. We redeemed the full reward balance as a statement credit. We then initiated a return through the merchant's standard process and received a refund on the original transaction. The refund posted to the card account promptly, but the reward ledger was not adjusted until the next statement close. At that point, Issuer~F correctly created a negative reward entry, debiting the ledger by the amount corresponding to the refunded transaction. The negative balance persisted: subsequent reward accruals first offset the negative balance before becoming redeemable again.

The critical observation is the \emph{window} between reward redemption and clawback. Because points are redeemable immediately upon settlement but clawback occurs only at statement close, a user who redeems rewards and then initiates a refund within the same billing cycle extracts value that will not be debited until the following statement close. Card issuers generally do not permit account closure while any outstanding obligation remains, so the negative balance cannot be escaped through account termination. However, a user who redeems, refunds, and then simply stops making qualifying purchases on the card can sustain a negative reward balance indefinitely, effectively retaining the redeemed value at zero net cost for an unbounded period. The system eventually converges toward Reward Integrity only if the user continues normal card activity.

\paragraph{Classification}
This behavior is an instance of V3 (Inconsistent Negative Balance Handling), specifically the V3(b) sub form where the timing asymmetry between instant reward availability and batched clawback creates an exploitable window. The \textsc{OnRefund} handler fires correctly and the negative balance entry is created on schedule; the workflow violation
occurs at the redemption gate, which does not enforce that
outstanding clawback has settled before permitting
redemption. This maps to CWE\nobreakdash-841 (Improper
Enforcement of Behavioral Workflow), as the required
sequence of clawback settlement prior to redemption
eligibility is not enforced~\cite{cwe841}. In the OWASP Business Logic Abuse Top~10, this aligns with Class~4 (Sequential State Bypass)~\cite{owaspbla2025}.

\paragraph{Impact}
The financial exposure per incident is smaller than in Cases~I and II because the negative entry does eventually offset future rewards during normal continued use. The exploit is not a permanent one time extraction (as in Cases~I and II where no clawback ever occurs) but rather a timing based extraction where the user obtains an interest free float of reward value for one or more billing cycles. The issuer bears the cost of this float and the risk of non recovery if the user reduces or ceases card activity.

\begin{takeawaybox}
\textbf{Case III-B Takeaway.}
Creating a negative reward balance is necessary but not sufficient. When reward availability is instant but clawback is batched to statement close, the timing asymmetry enables users to extract value before the adjustment occurs. Robust designs (Case~III-A, Issuer~C) eliminate this gap by processing clawbacks immediately upon refund settlement.
\end{takeawaybox}

\paragraph{Three Tier Summary}
Table~\ref{tab:issuer-comparison} summarizes the comparative behavior across all six issuers studied, ordered A through F. The results reveal a three tier spectrum: fully vulnerable (Issuers A and B, where refunds do not trigger cross cycle reward clawback), fully robust (Issuers C, D, and E, indefinite negative balance enforcement), and partially robust (Issuer~F, where rewards are available instantly but clawback is deferred to statement close, creating a timing based escape path).

\begin{table*}[t]
\small
\centering
\caption{Comparative issuer behavior across reward dimensions. $\checkmark$ = property satisfied; $\times$ = violated; $\sim$ = partially satisfied (conditional on continued card usage; balance persists indefinitely, but recovery depends on future qualifying spend which the user can unilaterally withhold).}
\label{tab:issuer-comparison}
\begin{tabular}{@{}lcccccc@{}}
\toprule
\textbf{Property} & \textbf{A} & \textbf{B} & \textbf{C} & \textbf{D} & \textbf{E} & \textbf{F} \\
\midrule
Reward Timing              & Instant  & Stmt.\ close & Instant & Stmt.\ close & Stmt.\ close & Instant \\
Refund Adjustment          & None     & None         & Immediate & Stmt.\ close & Stmt.\ close & Stmt.\ close \\
Neg.\ Balance Support      & N/A      & N/A          & Indefinite & Indefinite & Indefinite & Indefinite$^{*}$ \\
Reward Integrity           & $\times$ & $\times$     & $\checkmark$ & $\checkmark$ & $\checkmark$ & $\sim$ \\
Refund Reward Consistency  & $\times$ & $\times$     & $\checkmark$ & $\checkmark$ & $\checkmark$ & $\sim$ \\
\bottomrule
\multicolumn{7}{@{}l}{\scriptsize $^{*}$Balance persists indefinitely, but recovery depends on future qualifying spend, which the user can unilaterally withhold.}
\end{tabular}
\end{table*}

% ============================================================
% 6  FORMAL MODEL AND DEFENSIVE ALGORITHMS
% ============================================================
\section{Formal Model and Defensive Algorithms}
\label{sec:defensive}

We now present the formal state machine model and a suite of defensive pseudo algorithms. Our goal is not to prescribe a specific implementation, but to capture design patterns that enforce Reward Integrity and Refund Reward Consistency. While a full machine checked specification is beyond the scope of this paper, the four algorithms and three lemmas constitute a complete invariant preserving design pattern that practitioners can adapt to their specific execution environments.

\paragraph{Transaction Reward State Machine}
We model each transaction $t$ and its associated reward state as a pair $(t.\mathit{status}, R[t.\mathit{id}])$. The full system state consists of transaction states (\tsc{PENDING}, \tsc{SETTLED}, \tsc{PART\_REF}, \tsc{REFUNDED}, \tsc{CHARGEBACK}), a mutable per transaction reward record $R[t.\mathit{id}]$ (initially $0$, decremented on each partial clawback), an immutable original reward record $R_{\text{orig}}[t.\mathit{id}]$ (set once at settlement, never modified), a cumulative refund counter $\mathsf{TotalRefunded}[t.\mathit{id}]$ (initially $0$), and a user level reward ledger $B[U]$ that aggregates contributions from all transactions plus redemptions. All per transaction records ($R$, $R_{\text{orig}}$, $\mathsf{TotalRefunded}$) must persist for the lifetime of the account; garbage collecting these records after $N$ billing cycles would re-expose the system to the V1 vulnerability for late refunds. A well formed execution requires that every transition to \tsc{SETTLED} may trigger reward crediting, and every transition to \tsc{PART\_REF}, \tsc{REFUNDED}, or \tsc{CHARGEBACK} must trigger a compensating adjustment. In Case~I, this compensating step is entirely absent; in Case~II, it exists within the billing cycle but not across cycle boundaries.

\begin{figure}[t]
\centering
\resizebox{\columnwidth}{!}{%
\begin{tikzpicture}[
  node distance=2.0cm,
  every node/.style={font=\small},
  state/.style={draw, circle, minimum size=1.1cm, very thick, font=\scriptsize\bfseries},
  arr/.style={-{Stealth[length=2.2mm, width=1.8mm]}, very thick},
  lbl/.style={font=\tiny, midway, fill=white, inner sep=1pt, text=darkgray}
]
  \node[state, fill=softblue, draw=accentblue] (P) {\tsc{PND}};
  \node[state, fill=softgreen, draw=accentgreen, right=2.5cm of P] (S) {\tsc{SET}};
  \node[state, fill=softteal, draw=accentteal, below=1.5cm of S] (PR) {\tsc{P\_REF}};
  \node[state, fill=softamber, draw=accentamber, right=2.0cm of PR] (R) {\tsc{REF}};
  \node[state, fill=softred, draw=accentred, left=2.0cm of PR] (C) {\tsc{CHG}};

  \draw[arr, accentblue] (P) -- node[lbl, above] {settle / OnSettled} (S);
  \draw[arr, accentteal] (S) -- node[lbl, right, pos=0.4] {partial refund} (PR);
  \draw[arr, accentamber] (PR) -- node[lbl, below] {full refund} (R);
  \draw[arr, accentred] (S) -- node[lbl, left, pos=0.4] {dispute} (C);
  \draw[arr, darkgray, dashed] (P) to[bend right=40] node[lbl, below, pos=0.43] {cancel} (R);
  \draw[arr, accentteal] (PR) to[loop below, looseness=5, out=250, in=290] node[lbl, below] {partial refund} (PR);
\end{tikzpicture}
}%
\caption{Transaction reward state machine. Transitions to \tsc{SET} (Settled) trigger reward crediting via Algorithm~\ref{alg:settlement}. Transitions to \tsc{P\_REF} (Partially Refunded) or \tsc{REF} (Fully Refunded) trigger proportional clawback via Algorithm~\ref{alg:refund}. The self loop on \tsc{P\_REF} represents sequential partial refunds on the same transaction. The cancel transition from \tsc{PND} (Pending) does not involve rewards. The \tsc{SET}$\to$\tsc{CHG} transition triggers reward reversal atomically via Assumption~A5.}
\label{fig:statemachine}
\end{figure}
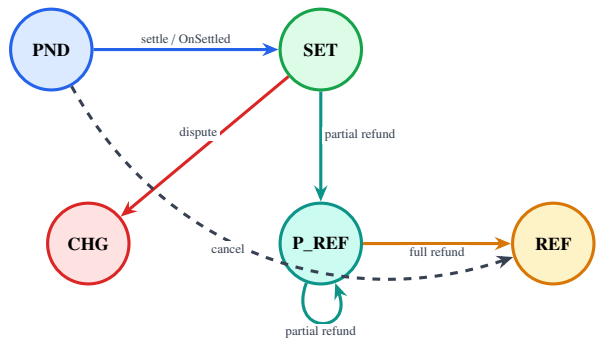

Figure~\ref{fig:statemachine} illustrates the state machine with required transitions and reward operations.

We now present four defensive algorithms. Together, they enforce both invariants across instant reward and statement cycle architectures. Algorithm~\ref{alg:settlement} handles reward crediting upon settlement. Algorithm~\ref{alg:refund} applies proportional clawback on refunds, directly closing the Case~I loophole. Algorithm~\ref{alg:canredeem} gates redemption against balance constraints. Algorithm~\ref{alg:reconcile} reconciles statement cycle systems against pending refunds, addressing Case~II.

\begin{algorithm}[t]
  \caption{RewardOnSettlement}
  \label{alg:settlement}
  \KwIn{Transaction $t = (\mathit{id}, u, m, a, c)$}
  \KwOut{Updated balance $B[u]$ and records $R[\mathit{id}]$, $R_{\text{orig}}[\mathit{id}]$}

  $\mathit{rate} \gets \mathsf{RewardRate}(c)$\;
  $\mathit{cap} \gets \mathsf{MonthlyCap}(u, c)$\;
  $\mathit{used} \gets \mathsf{MonthlyUsed}(u, c)$\;
  $r \gets \min(\mathit{rate} \cdot a,\; \mathit{cap} - \mathit{used})$\;
  \If{$r > 0$}{
    $B[u] \gets B[u] + r$\;
    $R[\mathit{id}] \gets r$\;
    $R_{\text{orig}}[\mathit{id}] \gets r$ \tcp*{immutable: never modified}
    $\mathsf{MonthlyUsed}[u, c] \gets \mathit{used} + r$\;
    $\mathsf{LogRewardEvent}(\mathit{id}, u, +r, \text{``settle''})$\;
  }
  $t.\mathit{status} \gets \tsc{SETTLED}$\;
\end{algorithm}

Algorithm~\ref{alg:settlement} computes the reward as the product of the category rate and transaction amount, capped by the remaining monthly allowance. The $\mathsf{MonthlyUsed}$ counter is incremented by the granted reward, ensuring that subsequent transactions in the same period observe the reduced cap headroom. Two per transaction records are stored: $R[\mathit{id}]$ tracks the current reward balance and is decremented on each partial refund, while $R_{\text{orig}}[\mathit{id}]$ preserves the original settlement reward as an immutable reference for proportional clawback computation. This separation is essential for correctness under sequential partial refunds, as discussed below. The status assignment on ($t.\mathit{status} \gets \tsc{SETTLED}$) is a no op when Algorithm~\ref{alg:settlement} is invoked by an event driven handler on already settled transactions; it is required when invoked from Algorithm~\ref{alg:reconcile}'s Phase~3 on \tsc{PENDING} transactions.

\begin{algorithm}[t]
  \caption{RewardOnRefund}
  \label{alg:refund}
  \KwIn{Transaction $t = (\mathit{id}, u, m, a, c)$, refund amount $x$}
  \KwOut{Updated balance $B[u]$ and record $R[\mathit{id}]$}

  \If{$t.\mathit{status} \notin \{\tsc{SETTLED}, \tsc{PART\_REF}\}$}{
    \Return \tcp*{skip non eligible transactions}
  }
  $x_{\text{prev}} \gets \mathsf{TotalRefunded}[\mathit{id}]$\;
  \If(\tcp*[f]{requires A3 serialization}){$x_{\text{prev}} + x > a$ \textbf{or} $x \le 0$}{
    \Return \tcp*{invalid: exceeds original amount}
  }

  $p \gets x / a$ \tcp*{fraction of original amount}
  $r_{\text{claw}} \gets p \cdot R_{\text{orig}}[\mathit{id}]$ \tcp*{use immutable original}

  $B[u] \gets B[u] - r_{\text{claw}}$ \tcp*{balance may go negative}
  $R[\mathit{id}] \gets R[\mathit{id}] - r_{\text{claw}}$\;
  $\mathsf{TotalRefunded}[\mathit{id}] \gets x_{\text{prev}} + x$\;
  \If{$t.n = \mathsf{CurPeriod}(u)$}{
    $\mathsf{MonthlyUsed}[u, t.c] \gets \max(0,\, \mathsf{MonthlyUsed}[u, t.c] - r_{\text{claw}})$\;
  }
  $\mathsf{LogRewardEvent}(\mathit{id}, u, {-r_{\text{claw}}}, \text{``refund''})$\;

  \eIf{$x_{\text{prev}} + x = a$}{
    $t.\mathit{status} \gets \tsc{REFUNDED}$\;
  }{
    $t.\mathit{status} \gets \tsc{PART\_REF}$\;
  }
\end{algorithm}

Algorithm~\ref{alg:refund} computes the clawback as the refund fraction $p = x/a$ times the \emph{immutable} original reward $R_{\text{orig}}[\mathit{id}]$ stored at settlement time. Using the original rather than the current (already decremented) reward value is essential for correctness under sequential partial refunds: if two partial refunds of \$40 each arrive on a \$100 transaction with an original \$5 reward, each must claw back $0.4 \times \$5 = \$2$, totaling \$4. Using the current $R[\mathit{id}]$ for the second refund would yield only $0.4 \times \$3 = \$1.20$, under clawing by \$0.80. The $\mathsf{MonthlyUsed}$ counter is decremented on lines~10--11 only when the refund falls within the same billing period as the original purchase, restoring cap headroom consumed by the original reward; for cross cycle refunds, the original period's cap counter has already been finalized, and decrementing the current period's counter would incorrectly create cap headroom that was never consumed in the current period. The negative balance created by the clawback is the primary defense against cross cycle abuse: even though the monthly cap resets in the new period, the user cannot extract the negative balance debt, which must be repaid from future earnings before any redemption is possible. The crucial design choice is that $B[u]$ is permitted to become negative, creating a debt that subsequent reward earnings must repay before the balance becomes redeemable. Note that Algorithm~\ref{alg:refund}'s entry guard accepts \tsc{SETTLED} and \tsc{PART\_REF} but not \tsc{CHARGEBACK}. This is deliberate: chargebacks are handled by the issuer's chargeback pipeline under Assumption~A5, which requires the same reward adjustment logic to be triggered through the dispute resolution callback rather than the standard refund path.

Table~\ref{tab:walkthrough} illustrates Algorithm~\ref{alg:refund}'s behavior on a concrete example with sequential partial refunds, demonstrating the role of $R_{\text{orig}}$ in maintaining correct proportional clawback.

\begin{table}[t]
\small
\centering
\caption{Walkthrough: \$100 purchase, 5\% rate, two sequential \$50 refunds. $R_{\text{orig}}$ is immutable; $R[\mathit{id}]$ tracks the running balance.}
\label{tab:walkthrough}
\begin{tabular}{@{}lcccc@{}}
\toprule
\textbf{Event} & $R_{\text{orig}}$ & $R[\mathit{id}]$ & $B[u]$ & \textbf{Status} \\
\midrule
Settle \$100  & \$5.00 & \$5.00 & +\$5.00 & \tsc{SET} \\
Refund \$50   & \$5.00 & \$2.50 & +\$2.50 & \tsc{P\_REF} \\
Refund \$50   & \$5.00 & \$0.00 & ~~\$0.00 & \tsc{REF} \\
\bottomrule
\end{tabular}
\end{table}

\begin{algorithm}[t]
  \caption{CanRedeem}
  \label{alg:canredeem}
  \KwIn{User $u$, requested redemption amount $y$}
  \KwOut{$\mathsf{true}$ if allowed, else $\mathsf{false}$}

  \If{$\mathsf{CurrentDate}() < \mathsf{RedemptionHoldUntil}[u]$}{
    \Return $\mathsf{false}$ \tcp*{grace period active}
  }
  \If{$B[u] - y < B_{\min}$}{
    \Return $\mathsf{false}$ \tcp*{insufficient balance}
  }
  \Return $\mathsf{true}$\;
\end{algorithm}

Algorithm~\ref{alg:canredeem} enforces two guards before permitting redemption. Let $B_{\min} \in \mathbb{R}$ be an issuer configured minimum balance threshold, typically $0$, below which no redemption is permitted; when $B_{\min} = 0$, no redemption is possible while $B[u] < 0$, providing complete debt enforcement under Assumption~A4. First, it checks whether the user is within an active grace period hold set by Algorithm~\ref{alg:reconcile}; this prevents premature redemption in statement cycle systems during the post close window when late refunds may still arrive. Second, it checks that the post redemption balance would not fall below $B_{\min}$, which is typically $0$. Together, these guards ensure that both the grace period defense and the negative balance defense are enforced at the redemption boundary. Algorithm~\ref{alg:canredeem} is a predicate only; the caller is responsible for executing the actual redemption (debiting $B[u] \gets B[u] - y$) upon receiving $\mathsf{true}$. For deployments that do not invoke Algorithm~\ref{alg:reconcile}, the grace period mechanism should be explicitly configured based on the reconciliation architecture: if no batch reconciliation component exists under any circumstances, the guard can be made permanently inactive; if any batch reconciliation is performed, even occasionally, the grace hold must remain active.

We define three query functions used by Algorithm~\ref{alg:reconcile}. $\mathsf{CurPeriod}(u)$ returns the billing period $n$ such that the current date falls within user $u$'s active billing cycle $[\mathsf{StmtOpen}(n), \mathsf{StmtClose}(n)]$. $\mathsf{TransactionsInPeriod}(u, n)$ returns all transactions whose billing period is $n$. $\mathsf{PendingRefunds}(u)$ returns the set of transaction identifiers with refunds that have arrived but not yet been incorporated into reward computations, restricted to current period transactions. $\mathsf{LateRefunds}(u, n)$ returns the set of $(\mathit{transaction}, \mathit{refund\_amount})$ pairs where the transaction's billing period is strictly less than $n$, the refund arrived during period $n$, and $\mathsf{TotalRefunded}[\mathit{id}] + x \le t.a$ (i.e., not yet fully processed). By construction, $\mathsf{PendingRefunds}$ covers current period transactions and $\mathsf{LateRefunds}$ covers prior period transactions, so the two sets are disjoint.

\begin{algorithm}[t]
  \caption{StatementCycleReconcile}
  \label{alg:reconcile}
  \KwIn{User $u$, billing period $n$}
  \KwOut{Adjusted reward set for period $n$}

  $\mathcal{T}_n \gets \mathsf{TransactionsInPeriod}(u, n)$\;
  $\mathcal{R}_{\text{pending}} \gets \mathsf{PendingRefunds}(u)$\;

  \ForEach{$t \in \mathcal{T}_n$}{
    \If{$t.\mathit{id} \in \mathcal{R}_{\text{pending}}$}{
      $x \gets \mathsf{RefundAmount}(t.\mathit{id})$\;
      $t.\mathit{eligible} \gets t.a - x$ \tcp*{local: original $t.a$ preserved}
    }
    \lElse{$t.\mathit{eligible} \gets t.a$}
  }

  \tcp{Retroactive: prior period txns with late refunds}
  $\mathcal{R}_{\text{late}} \gets \mathsf{LateRefunds}(u, n)$\;
  \ForEach{$(t, x) \in \mathcal{R}_{\text{late}}$}{
    \If{$t.\mathit{status} \in \{\tsc{SETTLED}, \tsc{PART\_REF}\}$}{
      $\mathsf{RewardOnRefund}(t, x)$ \tcp*{Alg.~\ref{alg:refund}}
    }
  }

  \ForEach{$t \in \mathcal{T}_n$ \textbf{where} $t.\mathit{eligible} > 0$ \textbf{and} $t.\mathit{status} = \tsc{PENDING}$}{
    $a_{\text{save}} \gets t.a$\;
    $t.a \gets t.\mathit{eligible}$ \tcp*{use refund adjusted amount}
    $\mathsf{RewardOnSettlement}(t)$ \tcp*{Alg.~\ref{alg:settlement}}
    $t.a \gets a_{\text{save}}$ \tcp*{restore original for future clawback}
  }

  $\mathsf{RedemptionHoldUntil}[u] \gets \mathsf{StmtCloseDate}(n) + \Delta_{\text{grace}}$\;
\end{algorithm}

Algorithm~\ref{alg:reconcile} addresses the Case~II timing gap for statement cycle systems. It operates in three phases. First, it cross references all period~$n$ transactions against pending refunds, computing a reward eligible amount $t.\mathit{eligible}$ for each transaction without mutating the original $t.a$ (which must be preserved for future clawback computations in Algorithm~\ref{alg:refund}). Second, it performs retroactive clawback on transactions from prior periods that received late arriving refunds during period~$n$, invoking Algorithm~\ref{alg:refund} for each; this closes the cross cycle gap that is the attacker's primary exploit path in Case~II, where returns are strategically timed to fall in a different billing cycle than the original purchase. Third, it computes rewards for remaining eligible transactions that are still in \tsc{PENDING} status; the status guard ensures that transactions already rewarded by a live event handler are not double credited (the \tsc{PENDING} check excludes all post Algorithm~\ref{alg:settlement} states: \tsc{SETTLED}, \tsc{PART\_REF}, \tsc{REFUNDED}, and \tsc{CHARGEBACK}, preventing conflicts in hybrid architectures where both live handlers and batch reconciliation coexist). The save/restore pattern temporarily substitutes $t.\mathit{eligible}$ for $t.a$ during the Algorithm~\ref{alg:settlement} call, so that $R_{\text{orig}}[\mathit{id}]$ reflects the refund adjusted eligible amount rather than the gross transaction amount. This is correct for statement cycle systems because rewards were never earned on the pending refunded portion; for event driven systems where Algorithm~\ref{alg:settlement} is called on the full $t.a$ at settlement time, $R_{\text{orig}}[\mathit{id}]$ reflects the gross amount. Both paths are consistent with Lemma~2 because Algorithm~\ref{alg:refund} always divides by the original $t.a$, producing the correct proportional clawback in either case. Finally, it sets the redemption hold. We note an important limitation: the grace period provides protection only for refunds that arrive within $\Delta_{\text{grace}}$ of statement close. For merchants with extended or open ended return windows (for example, 90 to 365 days, or unlimited as some warehouse retailers offer), a refund arriving months after the grace period has expired will not be intercepted by Algorithm~\ref{alg:reconcile}'s same period deduction phase. However, Algorithm~\ref{alg:reconcile}'s retroactive loop explicitly invokes Algorithm~\ref{alg:refund} for late arriving refunds, so the cross cycle gap is closed within one billing cycle of the refund's arrival, provided per transaction records have not been garbage collected. For refunds that arrive between statement closes (outside any reconciliation run), a production system with a live event handler would invoke Algorithm~\ref{alg:refund} directly upon refund posting. The grace period hold on redemption is therefore a complementary mechanism for statement cycle systems, not a substitute for the core per event clawback logic. For dormant accounts where multiple reconciliation cycles are skipped, the retroactive loop correctly processes all accumulated late refunds in the next active reconciliation run; the one billing cycle latency bound applies to accounts with regular card activity.

\paragraph{Correctness Argument}
We present a semi formal proof sketch that the proposed algorithms maintain both invariants. We use the term ``semi formal'' deliberately: a full machine checked proof would require a complete specification of the issuer's transaction processing pipeline, which is outside our scope. Instead, we state the key properties as lemmas and provide reasoning under stated assumptions.

\emph{Assumption A1:} Every refund event is eventually delivered to the reward engine (guaranteed by the honest payment network $N$). \emph{Assumption A2:} The reward engine processes events atomically (no partial updates). \emph{Assumption A3:} Concurrent refunds on the same transaction are serialized. \emph{Assumption A4 (Account Closure Constraint):} The issuer does not permit user initiated account closure while $B[u] < 0$; outstanding negative balances must be settled before account termination. Issuer initiated closure (fraud, program discontinuation) is outside the scope of our model and may require contractual or regulatory treatment. \emph{Assumption A5 (Chargeback Atomicity):} Upon receiving the payment network's dispute resolution outcome, the issuer's chargeback handling pipeline atomically updates both the transaction principal record and the associated reward records ($R[\mathit{id}]$, $\mathsf{TotalRefunded}[\mathit{id}]$, and $B[u]$). This is an issuer side implementation requirement, not a network provided guarantee; issuers deploying the proposed algorithms must ensure their chargeback callback triggers the same reward adjustment logic as a standard refund event.

\textbf{Lemma 1 (Per Transaction Bound).} After Algorithm~\ref{alg:settlement} executes on transaction $t$, $R[t.\mathit{id}] \le \mathsf{RewardRate}(t.c) \cdot t.a$ and $R[t.\mathit{id}] \le \mathsf{MonthlyCap}(u, t.c) - \mathsf{MonthlyUsed}(u, t.c)$. This follows directly from the $\min$.

\textbf{Lemma 2 (Proportional Clawback).} After Algorithm~\ref{alg:refund} processes a refund of amount $x$ on transaction $t$ with original amount $t.a$, the clawback amount equals $(x/t.a) \cdot R_{\text{orig}}[t.\mathit{id}]$, where $R_{\text{orig}}$ is the immutable reward stored at settlement time (for statement cycle systems using Algorithm~\ref{alg:reconcile}, this refers to the reconciliation time when the refund adjusted eligible amount is used, not the network settlement event). For a full refund ($x = t.a$), the entire original reward is clawed back. For same period refunds ($t.n = \mathsf{CurPeriod}(u)$), the $\mathsf{MonthlyUsed}$ counter is correspondingly decremented, restoring cap headroom. For cross cycle refunds, no cap adjustment occurs; clawback operates solely through the balance mechanism.

\textbf{Lemma 3 (Negative Balance and Grace Period Enforcement).} Algorithm~\ref{alg:canredeem} blocks redemption under two conditions: (a) if the current date falls within the grace period window $[\mathsf{StmtClose}(n),\; \mathsf{StmtClose}(n) + \Delta_{\text{grace}}]$ set by Algorithm~\ref{alg:reconcile}, and (b) if $B[u] - y < B_{\min}$. Guard~(a) prevents premature redemption during the configurable grace window; for merchants whose return windows extend beyond $\Delta_{\text{grace}}$, this guard provides partial but not complete protection, and Algorithm~\ref{alg:reconcile}'s retroactive $\mathsf{LateRefunds}$ loop provides the complementary clawback mechanism for refunds arriving in subsequent periods. Guard~(b) creates an automatic debt recovery mechanism: if $B[u] < 0$ after a clawback, no redemption is possible until future reward accruals restore the balance.

\textbf{Theorem (Reward Integrity).} Under assumptions A1--A5, at any point after all refund events delivered under A1 have been processed under A2, $\mathsf{NetReward}(U) \le f(\mathsf{NetSpend}(U), \mathsf{PromoRules})$. During the bounded delivery window guaranteed by A1, the invariant may be temporarily violated; the magnitude of any temporary violation is bounded by $\sum_{t \in \mathcal{P}} R_{\text{orig}}[t.\mathit{id}]$, where $\mathcal{P}$ is the set of transactions with delivered but unprocessed refund events. Assumption A1 further requires that per transaction records $R[\mathit{id}]$, $R_{\text{orig}}[\mathit{id}]$, and $\mathsf{TotalRefunded}[\mathit{id}]$ are not garbage collected before the refund arrives; as established in the state machine description, these records must persist for the lifetime of the account. \emph{Sketch:} By Lemma~1, each transaction contributes at most its rate bounded reward $R_{\text{orig}}[t.\mathit{id}] = \min(\mathsf{RewardRate}(t.c) \cdot t.a,\; \mathit{cap} - \mathit{used})$. By Lemma~2, each refund of amount $x_t$ removes $(x_t/t.a) \cdot R_{\text{orig}}[t.\mathit{id}]$. Summing over all transactions, the net reward equals $\sum_t R_{\text{orig}}[t.\mathit{id}] \cdot (1 - x_t/t.a)$. For a fully refunded transaction ($x_t = t.a$), this term equals zero, contributing no net reward; partially refunded transactions contribute proportionally. By Lemma~1, each $R_{\text{orig}}[t.\mathit{id}] \le \mathsf{RewardRate}(t.c) \cdot t.a$ (since $\min(\mathit{rate} \cdot a, \mathit{cap} - \mathit{used}) \le \mathit{rate} \cdot a$). Therefore $R_{\text{orig}}[t.\mathit{id}] \cdot (1 - x_t/t.a) \le \mathsf{RewardRate}(t.c) \cdot (t.a - x_t) \le f(t.a - x_t, \mathsf{PromoRules})$, since $f(\cdot)$ encodes both the rate and the cap. Summing over all transactions yields the bound. Note that when caps bind, the actual net reward may be strictly less than $\sum_t \mathsf{RewardRate}(t.c) \cdot (t.a - x_t)$ because $R_{\text{orig}}$ is capped at settlement time; the algorithms are therefore more conservative than a simple rate times net spend formula would suggest.

\textbf{Refund Reward Consistency.} For instant event driven systems with a live $\mathsf{OnRefund}$ handler, RRC is bounded by the refund delivery latency under Assumption~A1; once the event arrives, Algorithm~\ref{alg:refund} processes the clawback atomically under Assumption~A2. For statement cycle systems using Algorithm~\ref{alg:reconcile}, consistency is achieved in two stages: in phase~1, same period refunds are deducted synchronously within the reconciliation run ($\Delta = 0$); in phase~2, cross cycle refunds are retroactively clawed back via Algorithm~\ref{alg:refund}'s invocation in the $\mathsf{LateRefunds}$ loop, applied within one billing cycle of the refund's arrival ($\Delta \le$ one cycle length). For pure batch cycle architectures (no live handler), refunds arriving after $\Delta_{\text{grace}}$ but before the next statement close are intercepted by Algorithm~\ref{alg:reconcile}'s retroactive loop at the next reconciliation run. During the window between $\Delta_{\text{grace}}$ expiry and the next reconciliation, Algorithm~\ref{alg:canredeem}'s grace hold is no longer active, so a redemption request in this window would succeed; this residual window is an acknowledged limitation of the batch cycle design that event driven architectures do not share.

We note limitations of this argument. The $\mathsf{MonthlyUsed}$ decrement in Algorithm~\ref{alg:refund} is guarded by a period check: cap headroom is only restored when the refund falls within the same billing period as the original purchase. For cross cycle refunds, no cap adjustment is made; the original period's cap consumption stands, and the clawback operates solely through the balance mechanism ($B[u]$ goes negative). This design is safe because the negative balance debt prevents value extraction regardless of cap state, though it means a same period replacement purchase may not benefit from restored cap headroom if the refund arrives in a later cycle. Concurrent refunds near the monthly cap boundary introduce a race on $\mathsf{TotalRefunded}$ and $\mathsf{MonthlyUsed}$: Assumption~A3 requires serialization, which is marked in the pseudocode via the atomic annotation on Algorithm~\ref{alg:refund}'s guard but must be enforced by the production implementation through a mechanism such as a database row lock or transactional event queue. We note that A3 serializes concurrent refunds on the same transaction but does not cover concurrent refunds on different transactions for the same user. Assumption~A3's per transaction serialization fully protects $\mathsf{TotalRefunded}[\mathit{id}]$, since that counter is scoped to a single transaction; the benign direction argument below applies only to the cross transaction $\mathsf{MonthlyUsed}$ race, which A3 does not cover. In the latter case, the $\mathsf{MonthlyUsed}$ race is benign because each $r_{\text{claw}}$ is computed from a transaction specific $R_{\text{orig}}$ that cannot be inflated by the race, and a lost write results in over restriction (higher $\mathsf{MonthlyUsed}$) rather than under restriction. Rounding drift from floating point arithmetic in $r_{\text{claw}} = (x/a) \cdot R_{\text{orig}}[\mathit{id}]$ is partially mitigated by anchoring all clawback computations to the immutable $R_{\text{orig}}[\mathit{id}]$ value, which prevents compounding across sequential partial refunds; residual per event rounding error is bounded but not eliminated. A production implementation should clamp $R[\mathit{id}] \gets \max(0, R[\mathit{id}] - r_{\text{claw}})$ to prevent $R[\mathit{id}]$ from going negative due to accumulated rounding; the maximum per transaction rounding error is bounded by one unit of least precision times the number of partial refund events. A complete treatment of the interleaving semantics of concurrent events under relaxed serialization is left to future work.

We note two additional scope limitations. First, for transactions partially refunded across billing cycles, cap accounting for the earlier cycle is modified by same period partial refunds but not by subsequent cross cycle partial refunds; a replacement purchase in the original period may therefore benefit from artificially restored cap headroom on a transaction that is later fully refunded in a subsequent cycle. This is a residual design limitation that does not affect the Reward Integrity bound (which operates on balances, not caps) but may cause slight over crediting within a single period's cap window. Second, the algorithms assume each transaction belongs to a single category label $t.c$; multi category transactions (such as split receipt purchases spanning grocery and pharmacy) would require per category reward allocation at settlement time with Algorithm~\ref{alg:refund}'s proportional clawback applied per category component. We leave the formal treatment of multi category splits to future work.

Table~\ref{tab:alg-invariant} maps each algorithm to the invariants it enforces and the vulnerability classes it mitigates.

\begin{table}[t]
\small
\centering
\caption{Algorithm to invariant and vulnerability class mapping.}
\label{tab:alg-invariant}
\resizebox{\columnwidth}{!}{%
\begin{tabular}{@{}lll@{}}
\toprule
\textbf{Algorithm} & \textbf{Invariant} & \textbf{Mitigates} \\
\midrule
1: RewardOnSettle   & RI (caps)      & Per Txn Bound  \\
2: RewardOnRefund   & RI, RRC            & V1, V2         \\
3: CanRedeem        & RI, RRC            & V3 (partial$^{\dagger}$)  \\
4: StmtReconcile    & RI, RRC            & V1, V2         \\
\bottomrule
\multicolumn{3}{@{}l}{\scriptsize $^{\dagger}$Guard (a): post-close window; guard (b): negative balance under A4.}
\end{tabular}%
}
\end{table}

% ============================================================
% 7  DISCUSSION
% ============================================================
\section{Discussion}
\label{sec:discussion}

\subsection{Practical Deployment Considerations}

Implementing the proposed algorithms in existing reward engines raises several practical considerations. First, per transaction reward tracking (the $R[t.\mathit{id}]$ record) requires that the reward engine maintain a link between each reward credit and its originating transaction identifier. Systems that compute rewards in aggregate (for example, ``total grocery spend this month times reward rate'') without per transaction records cannot perform proportional clawback and would need schema changes. Second, negative reward balances create user experience challenges: consumers may be confused or frustrated by seeing a negative balance, and issuers need clear communication strategies. The robust issuers in Case~III-A demonstrate that this is manageable in practice, as their systems handle negative balances without apparent customer service escalation. The partially robust design of Issuer~F (Case~III-B) suggests that some issuers have chosen to bound negative balance enforcement, likely as a deliberate trade off between fraud prevention and customer experience. Third, the grace period in Algorithm~\ref{alg:reconcile} introduces a delay in reward availability, which may reduce the perceived value of ``instant'' rewards. Issuers must balance fraud prevention against competitive pressure to offer immediate gratification. We note that when Assumption~A4 holds (account closure blocked while $B[u] < 0$), the negative balance mechanism (guard~b in Algorithm~\ref{alg:canredeem}) alone is sufficient for eventual correctness: the user cannot escape the debt. The grace period hold (guard~a) is therefore a defense in depth measure that reduces the \emph{residual window} during which a batch cycle user could redeem rewards before the next reconciliation run processes a pending clawback. For event driven architectures with live handlers, guard~(a) is unnecessary, as Algorithm~\ref{alg:refund} processes clawbacks immediately. Issuers deploying pure batch cycle systems without live handlers benefit most from guard~(a).

\subsection{Financial Sensitivity Analysis}

Table~\ref{tab:impact} provides illustrative sensitivity estimates of annual reward leakage under the DDRA for various abuse fractions and population sizes.

\begin{table}[t]
\small
\centering
\caption{Illustrative annual reward leakage under DDRA (\$50 monthly cap shown; figures scale proportionally for programs with different cap structures). All figures in millions of USD. These are sensitivity bounds, not empirical measurements.}
\label{tab:impact}
\begin{tabular}{@{}lccc@{}}
\toprule
\textbf{Abuse fraction} $p$ & $|U|\!=\!$100K & $|U|\!=\!$1M & $|U|\!=\!$10M \\
\midrule
0.1\% & 0.06 & 0.6 & 6.0 \\
1.0\% & 0.6  & 6.0 & 60.0 \\
5.0\% & 3.0  & 30.0 & 300.0 \\
\bottomrule
\end{tabular}
\end{table}

We acknowledge that this model is a simple parametric projection ($p \times |U| \times \$600$/year) and that the critical parameters ($p$ and $|U|$) are not empirically estimated from our data. The \$50 monthly cap assumes optimal attacker behavior, which may overstate per user extraction in practice. The table is intended solely to illustrate the \emph{sensitivity} of financial exposure to abuse fraction and population size, not to provide a point estimate of actual losses or to claim that these figures represent real world leakage. For context, publicly available industry data suggests that major digital wallet providers serve tens of millions of active users, and industry reports estimate that loyalty and promotion fraud collectively accounts for one to three billion dollars in annual losses globally~\cite{li2025promoguardian}. Returns abuse rates in e commerce are estimated at 5 to 10 percent of transactions in some segments~\cite{zhang2023fraudulentreturns,frei2020returns}. We do not claim that the DDRA vulnerability contributes to these figures at the levels shown in Table~\ref{tab:impact}; we present the sensitivity analysis to motivate the economic relevance of the architectural flaws we identify.

\subsection{Tiered Defense Architecture}

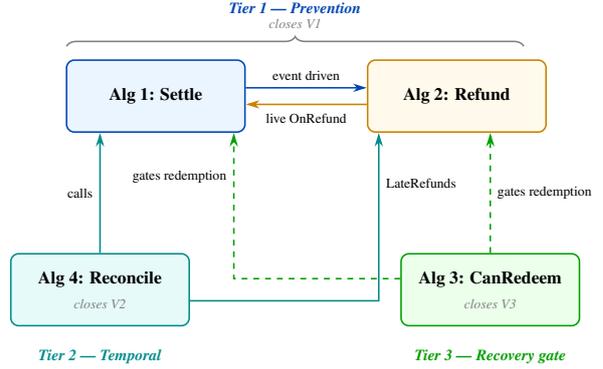
\begin{figure}[t]
\centering
\resizebox{\columnwidth}{!}{%
\begin{tikzpicture}[
    font=\sffamily,
    % Box styling: uniform size, bold text, rounded corners
    box/.style={
        draw, thick,
        rounded corners=4pt,
        minimum width=3.2cm,
        minimum height=1.3cm,
        align=center, 
        font=\small\bfseries
    },
    % Arrow styling: solid and dashed with filled triangles
    arr/.style={-{Stealth[length=2.5mm, width=1.5mm]}, thick},
    dashed_arr/.style={arr, dashed},
    % Labels styling
    lbl/.style={font=\scriptsize, text=black},
    tierlbl/.style={font=\footnotesize\bfseries\itshape},
    vulnlbl/.style={font=\scriptsize\itshape, text=gray}
]

    % --- Define Colors (Clean, Academic Palette) ---
    \definecolor{softblue}{RGB}{235, 245, 255}
    \definecolor{accentblue}{RGB}{0, 70, 180}
    \definecolor{softamber}{RGB}{255, 250, 235}
    \definecolor{accentamber}{RGB}{200, 130, 0}
    \definecolor{softteal}{RGB}{230, 250, 250}
    \definecolor{accentteal}{RGB}{0, 140, 140}
    \definecolor{softgreen}{RGB}{235, 255, 235}
    \definecolor{accentgreen}{RGB}{0, 160, 0}

    % --- 1. Place Boxes ---
    % We stagger the columns slightly so the vertical arrows can pass between the boxes 
    % without any line overlapping or touching a box it shouldn't.
    
    % Top Row (A1 and A2)
    \node[box, fill=softblue, draw=accentblue] (a1) at (0, 0) {Alg 1: Settle};
    \node[box, fill=softamber, draw=accentamber] (a2) at (5.4, 0) {Alg 2: Refund};
    
    % Bottom Row (A4 and A3) - Shifted outward to clear the routing channels
    \node[box, fill=softteal, draw=accentteal] (a4) at (-1.0, -3.5) {Alg 4: Reconcile\\[2pt] \textcolor{gray}{\normalfont\scriptsize\itshape closes V2}};
    \node[box, fill=softgreen, draw=accentgreen] (a3) at (6.0, -3.5) {Alg 3: CanRedeem\\[2pt] \textcolor{gray}{\normalfont\scriptsize\itshape closes V3}};

    % --- 2. Tier Labels ---
    \node[tierlbl, text=accentblue] at (2.5, 1.6) {Tier 1 — Prevention};
    \node[tierlbl, text=accentteal] at (-1.0, -4.7) {Tier 2 — Temporal};
    \node[tierlbl, text=accentgreen] at (6.0, -4.7) {Tier 3 — Recovery gate};

    % --- 3. Vulnerability Closure Annotation (closes V1) ---
    % Spans strictly from the left edge of A1 to the right edge of A2
    \draw[decorate, decoration={brace, amplitude=5pt}, draw=gray, thick] 
        (-1.6, 0.9) -- (6.6, 0.9) node[midway, above=0.2cm, vulnlbl] {closes V1};

    % --- 4. Draw SOLID Arrows ---
    % Arrow 1: A1 -> A2 (blue, upper gap)
    \draw[arr, accentblue] ([yshift=0.15cm]a1.east) -- node[above, lbl] {event driven} ([yshift=0.15cm]a2.west);

    % Arrow 2: A2 -> A1 (amber, lower gap)
    \draw[arr, accentamber] ([yshift=-0.15cm]a2.west) -- node[below=1pt, lbl] {live OnRefund} ([yshift=-0.15cm]a1.east);

    % Arrow 3: A4 -> A1 (teal, straight vertical up from C to A)
    \draw[arr, accentteal] (a4.north) -- node[left, lbl] {calls} (a4.north |- a1.south);

    % Arrow 4: A4 -> A2 (teal, L-shape: horizontal right then vertical up)
    % Exits slightly below center, turns up at x=3.6. Safely passes completely left of A3.
    \draw[arr, accentteal] ([yshift=-0.2cm]a4.east) -| node[right, pos=0.85, lbl] {LateRefunds} ([xshift=-1.4cm]a2.south);

    % --- 5. Draw DASHED Arrows ---
    % Arrow 5: A3 -> A2 (green dashed, straight vertical up from D to B)
    \draw[dashed_arr, accentgreen] (a3.north) -- node[right, lbl] {gates redemption} (a3.north |- a2.south);

    % Arrow 6: A3 -> A1 (green dashed, L-shape: horizontal left then vertical up)
    % Exits slightly above center, turns up at x=1.4. Safely passes completely right of A4.
    \draw[dashed_arr, accentgreen] ([yshift=0.2cm]a3.west) -| node[left, pos=0.85, lbl] {gates redemption} ([xshift=1.4cm]a1.south);

\end{tikzpicture}
}%

\caption{Algorithm interaction paths. In event-driven systems
(top row), settlement and refund events invoke Algorithms~1
and~2 directly as live handlers (Tier~1). In statement-cycle
systems, Algorithm~4 supplements this: Phase~1 deducts
same-period pending refunds before reward computation;
Phase~2 applies retroactive cross-cycle clawback via
$\mathsf{LateRefunds}$ by calling Algorithm~2; and Phase~3
settles rewards for transactions still \textsc{Pending} at
statement close by calling Algorithm~1 (Tier~2). Algorithm~3
acts as a cross-cutting redemption gate, enforcing both the
grace-period hold (Tier~2, guard~a) and negative-balance
enforcement (Tier~3, guard~b).}
\label{fig:algflow}
\end{figure}

Our defensive algorithms naturally organize into a three tier defense model that mirrors the three tier spectrum of issuer behavior observed in our case studies. Figure~\ref{fig:algflow} illustrates the two primary invocation paths and the cross cutting redemption gate.

\paragraph{Tier 1: Prevention (Algorithms 1 + 2)} Per transaction reward tracking with immutable $R_{\text{orig}}$ records and proportional clawback on every refund event eliminates V1 (Refund Insensitive Rewards, Case~I) entirely and provides the per transaction linkage that V2 systems lack. This tier requires no timing coordination, no grace periods, and no batch reconciliation; it operates on a per event basis. The robust issuers in Case~III-A (C, D, and E) implement this tier, which is why they are invulnerable to the DDRA pattern regardless of when refunds arrive.

\paragraph{Tier 2: Temporal Enforcement (Algorithms 3 + 4)} For statement cycle architectures that batch reward computation, Tier~1 alone is insufficient because a timing window exists between reward computation and refund arrival. Algorithm~\ref{alg:reconcile}'s three phase reconciliation (same period deduction, retroactive clawback via $\mathsf{LateRefunds}$, and reward computation with the \tsc{PENDING} guard) closes the Case~II timing gap. Algorithm~\ref{alg:canredeem}'s grace period hold provides additional protection during the $\Delta_{\text{grace}}$ window. This tier mitigates V2 (Temporally Scoped Adjustment) for batch architectures.

\paragraph{Tier 3: Recovery (Negative Balance + A4)} Even when Tiers~1 and~2 are imperfect, permitting $B[u]$ to go negative and enforcing Assumption~A4 (no account closure while $B[u] < 0$) provides eventual consistency. The user cannot extract the negative balance debt, and future reward earnings must first repay it. This tier addresses V3 (Inconsistent Negative Balance Handling) and provides a safety net for residual windows that Tiers~1 and~2 do not fully cover, such as the batch cycle window between $\Delta_{\text{grace}}$ expiry and the next reconciliation run.

\paragraph{Complementary Detection} Our prevention focused tiers complement detection based approaches such as PromoGuardian~\cite{li2025promoguardian} and behavioral fraud analysis~\cite{chu2023spatiotemporal,wang2022finegrained}, which identify suspicious patterns after the fact. Prevention provides a stronger foundation because: (i) detection systems can have false negatives, especially for honest but opportunistic users whose individual behavior closely resembles legitimate returns; (ii) even perfect detection cannot recover rewards that have already been redeemed and transferred out; and (iii) correct by construction logic reduces operational burden. Detection remains valuable as a fourth layer, particularly for identifying coordinated abuse at scale.

\subsection{Regulatory Considerations}

The Consumer Financial Protection Bureau (CFPB) in the United States has shown increasing interest in the transparency and fairness of credit card reward programs, including enforcement actions related to unfair, deceptive, or abusive acts or practices (UDAAP) in rewards program administration~\cite{cfpbrewards2024}. Our work highlights a related concern: when reward engines fail to enforce basic integrity invariants, the resulting inconsistencies can affect not only the issuer's financial exposure but also consumer trust and regulatory compliance. Issuers that implement robust clawback mechanisms (as in Case~III) are better positioned to demonstrate fair and consistent treatment.

\subsection{Generalization Beyond Cashback}

While our case studies focus on cashback and category based reward programs, we conjecture that the underlying vulnerability patterns may generalize to other reward modalities that face similar architectural challenges. Airline mileage programs that credit miles upon ticket purchase, hotel loyalty programs that grant points at check in, and cryptocurrency exchange platforms that offer trading fee rebates all must reconcile incentive ledgers against transaction reversals. Whether these systems exhibit the same V1, V2, or V3 patterns is an empirical question that we leave to future investigation.

% ============================================================
% 8  ETHICS AND DISCLOSURE
% ============================================================
\section{Ethics, Disclosure, and Methodology}
\label{sec:ethics}

\subsection{Experimental Methodology}

Our empirical observations are based on personal accounts at a small number of issuers that we legitimately hold, low value purchases (under \$50 per transaction) using standard refund processes solely to confirm reward behavior, and public documentation (terms and conditions, FAQ pages) of major reward and loyalty programs. We did not create fake identities or synthetic accounts, did not attempt large scale or automated exploitation, and did not interact with systems outside of published user interfaces.

\subsection{Responsible Disclosure}

All experiments used only accounts we personally and legitimately hold, involved low value transactions consistent with normal consumer use, and were designed to confirm architectural behaviors rather than to extract financial gain.

\paragraph{Case~I (Issuer~A)} We filed a report through the provider's private bug bounty program on October~25, 2025, including a vulnerability description, steps to reproduce the observed behavior, and a recommendation to implement proportional reward clawback on refund events. The provider acknowledged receipt on October~28, 2025, and confirmed the finding as a known vulnerability on January~14, 2026. More than 150~days have elapsed since our initial report, exceeding the standard 90~day coordinated disclosure window. We anonymize the provider and omit implementation details per the program's terms.

\paragraph{Case~II (Issuer~B)} We made sustained good faith efforts to locate a disclosure channel. We searched the provider's public website, security pages, support documentation, and legal notices for a vulnerability disclosure policy, security contact, or bug bounty program; none were found. On April~1, 2026, we contacted customer support by telephone and requested referral to a security or compliance team; this produced no usable contact. Later, we sent disclosure emails to the provider's \texttt{security@} and \texttt{abuse@} addresses. As of writing, no response has been received through any channel. In the absence of any discoverable disclosure channel after
exhausting reasonable direct contact avenues, we proceed with
anonymized publication, consistent with responsible disclosure
practice for vendors with no accessible vulnerability
reporting pathway~\cite{sun2021concession, wang2011shopfree}. The finding is a systemic design pattern (statement cycle batching without cross cycle refund reconciliation) described at the architectural level only, with all identifying details anonymized.
% Dated records of all attempts are retained. We acknowledge that our initial notification and this submission occurred in close temporal proximity, which does not satisfy a standard coordinated disclosure waiting period; we commit to providing Issuer~B an additional notification window before any camera ready publication and will inform the editor if contact is established.

\paragraph{Case~III-B (Issuer~F)} Unlike Cases~I and II, Issuer~F's reward engine does implement a refund adjustment mechanism: a proportional negative reward entry is created at statement close. The vulnerability we document is a timing asymmetry between instant reward availability at settlement and batched clawback at statement close, not a complete absence of refund handling. Because recovery occurs automatically during normal continued card usage, we characterize this as a design trade off with security implications rather than an unambiguous exploitable vulnerability. Accordingly, we did not pursue formal vulnerability disclosure for this case. We describe the behavior at the architectural level only, anonymize the provider throughout, and present it as a partial robustness pattern to inform secure reward engine design.

\subsection{Limitations}

Our study has several limitations that we acknowledge transparently. Most significantly, our empirical scope is small: we examine six issuers, constrained by the accounts we personally hold. This scope reflects a constraint intrinsic to the research domain. Opening a credit or debit card account in the United States requires government issued identity documentation, a Social Security Number, a credit inquiry, and issuer approval under applicable KYC and BSA/AML regulations; it is neither legally nor practically feasible for academic researchers to create large numbers of financial accounts for experimental purposes. This constraint is consistent with our adversary model, which characterizes the threat actor as a single consumer using one legitimate account per issuer. A study of six accounts across six distinct issuers is sufficient to instantiate the vulnerability classes we define, confirm the behavioral spectrum from fully vulnerable to fully robust, and motivate the defensive algorithms we propose. We make no population level statistical claims about vulnerability prevalence, and our economic impact estimates in Table~\ref{tab:impact} are explicitly labeled as illustrative sensitivity bounds rather than empirical measurements.

We lack internal visibility into backend systems, risk models, or fraud detection infrastructure at any issuer. We do not build a scanner or automated tool to infer reward logic at scale. The formal model and correctness argument are semi formal; a full machine checked proof would require specifying the complete event interleaving semantics, which we leave to future work. The grace period parameter $\Delta_{\text{grace}}$ in Algorithm~\ref{alg:reconcile} (suggested as 7 to 14 days) is an issuer configurable parameter that would require empirical calibration against actual refund arrival distributions, which we do not have access to. A well calibrated $\Delta_{\text{grace}}$ should cover the tail of the refund arrival distribution in the days immediately following statement close; for a 30 day merchant return window on a 30 day billing cycle, refunds arriving in the subsequent period up to 14 days post close represent the bulk of the exploitable window, though this varies by merchant and product category. Despite these limitations, we believe the combination of real world evidence (including a vendor acknowledged vulnerability), formal modeling, and actionable defensive algorithms provides sufficient value for the security engineering community.

% ============================================================
% 9  RELATED WORK
% ============================================================
\section{Related Work}
\label{sec:related}

Our work intersects research from web and systems security, operations and marketing science, and industry studies on promotion and returns abuse. To our knowledge, no prior work explicitly models card and wallet reward engines as state machines with refund aware integrity invariants, nor analyzes cashback specific reward abuse patterns through empirical case studies from production financial systems.

\paragraph{Business Logic Vulnerabilities}
Felmetsger et al.\ proposed automated detection of logic vulnerabilities by inferring expected workflows~\cite{felmetsger2010logic}. Pellegrino and Balzarotti demonstrated black box discovery of logic flaws from network traces~\cite{pellegrino2014blackbox}. Nabi argued for protecting business application logic integrity in e commerce systems through security focused frameworks~\cite{nabi2005secure,nabi2011framework}. Yu et al.\ propose a framework for modeling and analyzing logic vulnerabilities within e-commerce systems during the design phase~\cite{yu2023modeling}. Chen et al.\ showed that documentation and implementation inconsistencies in payment syndication services produce exploitable behavior~\cite{chen2019devils}. Ghorbansadeh and Shahriari proposed detecting logic vulnerabilities by comparing design with implementation~\cite{ghorbansadeh2020detecting}. Sun et al.\ propose GPTScan, a framework that detects logic vulnerabilities by combining large language models with traditional program analysis~\cite{sun2024gptscan}. The 2025 OWASP Business Logic Abuse Top~10 categorizes 63 vulnerability classes validated against over 76,000 CVEs~\cite{owaspbla2025}. These works motivate our state machine model, but they do not consider reward specific semantics.

\paragraph{Payment and E Commerce Security}
Wang et al.\ showed that integration errors in online payment flows enable attackers to obtain goods without paying, and similar ``infinite money'' logic flaws have been documented in web application security contexts where store credits or gift cards can be cycled without limits~\cite{wang2011shopfree}. Sun et al.\ analyzed ``Concession Abuse as a Service,'' documenting industrialized exploitation of customer service concessions~\cite{sun2021concession}. Van Goethem et al.\ demonstrated timeless timing attacks that exploit concurrency to leak secrets over remote connections~\cite{vangoethem2020timeless}; while the time scale differs by orders of magnitude, both our V2 class and timing attacks share the structural property that a security relevant event falls outside the system's processing window. Behavioral analysis for online payment fraud detection has been studied including spatial temporal pattern exploitation~\cite{chu2023spatiotemporal}, fine grained co-occurrence modeling for payment services~\cite{wang2022finegrained}, and hidden Markov models for credit card fraud~\cite{srivastava2008creditcard}. Du et al.\ studied app distribution fraud involving promotion investment manipulation~\cite{du2022appdistribution}. Bitaab et al.\ proposed large scale detection of fraudulent e commerce websites, identifying eight distinct types of non phishing financial fraud~\cite{bitaab2023beyondphish}; their follow up work uncovered coordinated campaigns of fraudulent shopping websites operating at scale~\cite{bitaab2025scammagnifier}. These detection oriented approaches highlight the gap between identifying fraud after the fact and the prevention by design methodology we propose: our algorithms ensure that reward logic is correct by construction rather than relying on post hoc anomaly detection. Our Case~I can be viewed as a cashback specific variant of the concession abuse patterns: the mis modeled artifact is the reward itself, which is not properly tied to the lifecycle of the underlying transaction.

\paragraph{Returns Abuse and Promotion Fraud}
Harris documented consumer exploitation of liberal return policies~\cite{harris2010fraudulent}. Frei et al.\ showed that product returns represent a growing problem with significant financial impact~\cite{frei2020returns,frei2023covidreturns}. Akturk et al.\ modeled return countermeasures~\cite{akturk2021managing}. Ketzenberg et al.\ applied analytics to segment return behavior~\cite{ketzenberg2020returns}. Liu and Du studied return policy abuse in e commerce~\cite{liu2023returnpolicy}. Zhang et al.\ provided a multichannel returns fraud framework~\cite{zhang2023fraudulentreturns}. Merlano et al.\ examined consumer perceptions of anti fraud interventions~\cite{merlano2024fraudulentreturns}. Von Zahn et al.\ proposed machine learning nudges to reduce returns~\cite{vonzahn2022smartnudge}. These works treat rewards as part of aggregate economics rather than as a security critical state machine.

\paragraph{Promotion Abuse Detection}
Vieira et al.\ examine how consumer program loyalty mediates the financial benefits that retailers derive from cashback strategies~\cite{vieira2022cashback}. Leveraging heterogeneous graph neural networks and hierarchical attention mechanisms, Ghosh et al.\ propose the GoSage framework specifically designed for incentive abuse detection~\cite{ghosh2023gosage}. Utilizing a risk-scoring methodology, Aprisadianti and Dwiyanti develop a novel application specifically designed to detect promotion abuse fraud~\cite{aprisadianti2023promotion}. Li et al.\ introduced PromoGuardian, a graph neural network for promotion fraud detection achieving 93 percent precision in production at Meituan~\cite{li2025promoguardian}. These detection systems complement our prevention focused approach: they identify abusive patterns after the fact, while we propose logic that prevents the abuse from being possible in the first place.

\paragraph{Cashback Economics}
Taleizadeh et al.\ showed that cashback strategies interact with refund policies and credit terms~\cite{taleizadeh2023cashback,taleizadeh2024pricing}. Wu et al.\ examined cashback tied to online reviews~\cite{wu2023praise}. These models assume compliant consumers and do not explore adversarial exploitation of reward accounting logic.

Our work is distinguished from adjacent fraud categories by a specific combination of characteristics: the attacker uses a legitimately held account, conducts real purchases through normal merchant channels, makes no false claims about product condition or non receipt, does not collude with other parties, and exploits a logic flaw in the reward engine rather than a technical vulnerability. Account takeover, synthetic identity fraud, and friendly fraud each lack one or more of these properties, and their mitigations (stronger authentication, identity verification, claim investigation) do not address the reward engine logic flaws we study.

% ============================================================
% 10  CONCLUSION
% ============================================================
\section{Conclusion}
\label{sec:conclusion}

% Note: Ensure you have these libraries loaded in your preamble:
% \usetikzlibrary{positioning, arrows.meta, calc, decorations.pathreplacing}
The vulnerabilities documented in this paper require no sophisticated tools or insider access; they are exploitable by any consumer who understands the basic mechanics of purchasing, redeeming rewards, and returning merchandise. Yet the defenses are equally straightforward: per transaction reward tracking, proportional clawback triggered by every refund event, and negative balance enforcement that persists until recovered. At least three of the six issuers we studied already deploy these patterns successfully, suggesting that robust reward logic is technically feasible even if we cannot claim from our small sample that it is yet industry wide practice. The key differentiator between a vulnerable system and a robust one is not the sophistication of fraud detection analytics but rather whether the reward engine treats refund events as first class state transitions that must update the reward ledger.

Through controlled experiments on six production issuer accounts, we documented a three tier spectrum of cashback reward handling on refunds, formalized the underlying flaws as violations of Reward Integrity and Refund Reward Consistency within a state machine model, and proposed defensive algorithms with a structured correctness argument. We believe this work makes a case for treating cashback reward engines with the same engineering rigor that the payments industry applies to authorization and settlement logic. As reward programs continue to grow in financial significance, the integrity of the incentive ledger deserves a place alongside the integrity of the payment ledger in security reviews, compliance assessments, and system design.

% ============================================================
% REFERENCES
% ============================================================

\end{document}